\documentclass[]{aastex631}

\usepackage{amsmath}
\usepackage{subfigure}
\usepackage{natbib}
\usepackage{todonotes}
\usepackage{booktabs}

\usepackage{hyperref}
\hypersetup{
    colorlinks=true,
    linkcolor=blue,
    filecolor=magenta,      
    urlcolor=blue,
    pdftitle={Overleaf Example},
    pdfpagemode=FullScreen,
    }
\usepackage{dcolumn}
\usepackage{bm}
\usepackage{orcidlink}
\usepackage{graphicx}

\begin{document}

\title[Cosmic Genealogy of SMBHBs across redshifts]{The Route to Unveil the Cosmic Genealogy of Supermassive Black Hole Binaries Using Nano-Hertz Gravitational Waves and Galaxy Surveys}

\author[0009-0005-9881-1788]{Mohit Raj Sah}
\affiliation{Department of Astronomy and Astrophysics, Tata Institute of Fundamental Research, Mumbai 400005, India}
\email{mohit.sah@tifr.res.in}

\author[0000-0002-3373-5236]{Suvodip Mukherjee}
\affiliation{Department of Astronomy and Astrophysics, Tata Institute of Fundamental Research, Mumbai 400005, India}
\email{suvodip@tifr.res.in}

\begin{abstract}

The nano-hertz (nHz) stochastic gravitational wave background (SGWB), generated by unresolved supermassive black hole binaries (SMBHBs), provides a unique probe of their population and its cosmic evolution. In this work, we explore the potential of uncovering the SMBHB population and its redshift dependence by combining the SGWB signal and its anisotropies with galaxy distribution through cross-correlation analyses. Using a Fisher analysis technique, we show that the SGWB power spectrum alone can not provide any information on the evolutionary history of SMBHBs, whereas the inclusion of the angular power spectrum of the SGWB and its cross-correlation with the galaxy distribution substantially improves constraints on the redshift evolution parameters. Assuming pulsar timing array (PTA) configurations achievable in the Square Kilometre Array (SKA) era, we find that the combined use of isotropic and anisotropic SGWB signals, together with galaxy surveys, can provide valuable measurements of the redshift evolution of the SMBH–galaxy connection and the frequency distribution of SMBHBs. These results highlight the potential of joint GW–galaxy studies to address the long-standing open question of SMBH growth and evolution across cosmic time.

\end{abstract}


\section{Introduction}

Pulsar Timing Arrays (PTAs) have opened a new observational window for studying nano-hertz (nHz) gravitational waves (GWs), whose primary source is believed to be the inspiraling supermassive black hole binaries (SMBHBs). By accurately measuring the arrival times of radio pulses from millisecond pulsars, PTA collaborations worldwide, which include the European PTA (EPTA; \cite{desvignes2016high}), the North American Observatory for Gravitational Waves (NANOGrav; \cite{mclaughlin2013north}), the Parkes PTA (PPTA; \cite{manchester2013parkes}), the Indian PTA (InPTA; \cite{joshi2018precision}), the Chinese PTA (CPTA; \cite{xu2023searching}), and MeerKAT PTA (MPTA; \cite{miles2023meerkat}) are actively searching for the nanohertz (nHz) stochastic gravitational wave background (SGWB). Their efforts have led to compelling evidence of nHz SGWB \citep{agazie2023nanograv,antoniadis2023second,zic2023parkes,xu2023searching,miles2025meerkat}.

The formation and evolution of SMBHs remain a long-standing question in astrophysics \citep{volonteri2007evolution,volonteri2010formation,latif2016formation,volonteri2021origins}. The study of nHz GWs offers a unique opportunity to investigate their population and cosmic evolution \citep{feng2020supermassive,padmanabhan2023unraveling,agazie2023nanogravSMBHB,sah2024imprints,sah2024discovering,sato2025evolution}. SMBHBs are believed to be a byproduct of galaxy mergers, where the gravitational interactions between merging galaxies bring their central SMBHs closer together, eventually forming a binary system. As these binaries lose energy through mechanisms such as dynamical friction, stellar scattering, and viscous drag, they eventually enter the GW decay regime, producing GW signals detectable in the PTA band \citep{sampson2015constraining,kelley2017massive,chen2017efficient,taylor2017constraints,izquierdo2022massive}. The study of these GWs provides a direct probe of the SMBHB population, shedding light on their mass distribution, merger rates, and correlations with the properties of their host galaxies. Moreover, such investigations enable a deeper understanding of the cosmic evolution of SMBHBs across redshift \citep{sah2024imprints,sah2024discovering}.

The SGWB is characterized by its frequency-dependent energy density spectrum (represented by $\Omega_{\rm GW}(f)$). This spectrum can provide valuable information about the sources of the SGWB.  However, this spectrum alone is insufficient to constrain the wide range of properties associated with the SMBHB population. To obtain deeper insights, the SGWB must be further analyzed through its anisotropies. The anisotropy in the SGWB signal is typically quantified using the auto-angular power spectrum, $C_{\ell}^{\rm GWGW}$, which characterizes the spatial distribution of the GW background \citep{mingarelli2013characterizing,taylor2013searching,taylor2020bright,agazie2023nanograv,sato2024exploring,raidal2024statistics}. However, the anisotropy is likely to be non-Gaussian, and $C_{\ell}^{\rm GWGW}$ is expected to be dominated by shot noise \citep{sah2024imprints}. {\cite{sah2025multi} proposed a novel technique for detecting the anisotropic SGWB, in which the anisotropic SGWB is directly stacked in regions of galaxy overdensity using a galaxy catalog as a tracer.} Recently, \cite{sah2024discovering} demonstrated the use of the cross-angular power spectrum ($C_{\ell}^{\rm gGW}$ ) between the nHz SGWB and galaxy catalogs. It was shown that the $C_{\ell}^{\rm gGW}$ carries the imprint of the cosmic evolution of the SMBH population. By leveraging galaxy survey data, such as those from the Rubin LSST Observatory \citep{ivezic2019lsst}, we can probe the evolution of the SMBHB population and their host galaxies over cosmic time \citep{sah2024discovering}.

In this work, we aim to assess the feasibility of using the SGWB power spectrum combined with its anisotropy to constrain the properties and evolution of the SMBHB population. By utilizing different configurations of future PTAs in combination with next-generation galaxy surveys, we evaluate the potential to extract robust constraints on the SMBHB population. We found that the SGWB power spectrum alone is inadequate for constraining SMBHB properties due to parameter degeneracies. However, incorporating the information from anisotropy in the signal significantly improves parameter estimation by breaking these degeneracies, enabling a more detailed characterization of the SMBHB population. 

The paper is organized as follows: in Sec. \ref{sec:SGWB_pop}, we discuss the SMBHB population model as well as the SGWB generated by the population; in Sec. \ref{sec:Aniso}, we discuss the angular power spectrum of the SGWB signal and its cross-correlation with the galaxy distribution; in Sec. \ref{sec:Map}, we explore how population parameters influence different components of the SGWB signal; in Sec. \ref{sec:Cross}, we review the cross-correlation of PTA timing residuals and the various noise components affecting these measurements; in Sec. \ref{sec:Fisher}, we present the result of the Fisher analysis.  Finally, in Sec. \ref{sec:Conc}, we discuss the conclusions and future prospects.

\section{GW Background \lowercase{{\large from}} SMBHB population}\label{sec:SGWB_pop}

The SGWB density from a population of SMBHB in unit solid angle can be written as \citep{phinney2001practical,sesana2008stochastic,christensen2018stochastic}

\begin{equation}
    \begin{aligned}
     \Omega_{\rm GW}(f, \hat n) = \frac{1}{4\pi\rho_{\rm c} c^2} \int dV & \int \prod\limits_{i}^{n} d\theta_{i}  \Big[\kappa(\Theta_{\rm n},f_r|{\rm z},\hat{n}) n_{\rm v}({\rm z},\hat{n})\Big] \Big[ \frac{1}{4 \pi d_{\ell}^{2} c} \frac{ dE_{\rm{gw}}(f_r,\Theta_{\rm n})}{{ dt_{\rm r}}}\Big],
    \end{aligned}
    \label{SGWB2}
\end{equation}
where  $\Theta_{\rm n}$ = $\{\theta_i\}_{i=1}^{\rm n}$ denotes the set of n GW source parameters, $f_r$ is the source frame frequency, $d_{\ell}$ is the luminosity distance, $n_{\rm v}({\rm z},\hat{n})$ represents the number of galaxies per unit comoving volume per unit solid angle, and $\frac{ dE_{\rm{gw}}(f_r, \Theta_{\rm n})}{{ dt_{\rm r}}}$ is the GW luminosity of a binary with source property $\Theta_n$. Finally, $\kappa(\Theta_{\rm n}, f_r \mid{\rm z},\hat{n})$ denotes the mean number of SMBHBs per galaxy, at redshift ${\rm z} $, with source property $\Theta_n$, emitting in a unit logarithmic frequency interval in $ f_r $. $\kappa(\Theta_{\rm n}, f_r \mid {\rm z},\hat{n})$ averaged over the sky is modeled as

\begin{equation}
    \begin{aligned}
        \overline{\kappa}( M_{\rm BH}, q, f_r|{\rm z}) =~ & \mathcal{N}~ (1+{\rm z})^{\xi} \int dM_{*} ~ P(M_{*}|{\rm z}) ~  P( M_{\rm BH}, q, f_r|M_{*}, {\rm z}),\\
        =~ &\mathcal{N}~(1+{\rm z})^{\xi} \int dM_{*} ~ P(M_{*}|{\rm z}) P( M_{\rm BH}| M_*,{\rm z}),\\
        & P(q| M_*,{\rm z})~ [P(f_r| M_*,{\rm z})~f_r],
    \end{aligned}
    \label{pop}
\end{equation}
where $M_{\rm BH}$  denotes the mass of the primary BH, $q$ is the mass ratio of the binary system, and $M_{*}$ is the stellar mass of the galaxy. $\mathcal{N}$ is a normalization constant quantifying the occupation fraction of SMBHBs in galaxies, while the factor $(1+{\rm z})^{\xi}$ parametrizes its redshift evolution. $P(M_{*}|{\rm z})$ is the stellar mass function of the galaxy, and  
$P(M_{\rm BH} | M_*, {\rm z})$ is modeled as a Gaussian distribution of the logarithm of the SMBH mass, conditioned on the stellar mass of the host galaxy.

\begin{equation}
    P( M_{\rm BH}| M_*,{\rm z}) \propto \frac{1}{M_{\rm BH}} \times \mathbf{exp}\Big(-\frac{(\mathrm{Log}_{10}[M_{\rm BH}] - \mathrm{Log}_{10}[\tilde{M}_{\rm BH}(M_{*})])^{2}}{2 \sigma_{m}^{2}}\Big),
\end{equation}
where 
\begin{equation}
    \mathrm{Log}_{10}(\tilde{M}_{\rm BH}) = \eta + \rho ~ \mathrm{Log}_{10}( M_{*}/10^{11} M_{\odot}) +\nu ~ {\rm z} , 
    \label{MBH1}
\end{equation}

with $\eta$, $\rho$, and $\nu$ being free parameters governing the scaling relation. The probability distributions for the mass ratio, $P(q | M_{*}, {\rm z})$ is modeled as

\begin{equation}
     P(q|  M_{*},{\rm z}) \propto \bigg\{
    \begin{array}{cl}
    & 1/q, \quad  0.01 < q < 1,\\
    & 0, ~~ \rm{else}. 
    \end{array}
    \label{q}
\end{equation}
For the entire analysis in this paper, we adopt the simplifying assumption of circular orbits. Consequently, each binary system emits GW monochromatically. Under this assumption, we model the source-frame frequency distribution, $P(f_r | M_{*}, {\rm z})$, as
\begin{equation}
    P( f_r|  M_{*},{\rm z}) \propto  f_{\rm r}^{\alpha ~+ ~ {\rm {\rm z}}  ~{\lambda}}, 
    \label{freq_dist}
\end{equation}
where the parameters $\alpha$, and $\lambda$ govern the frequency distribution and its redshift evolution. Astrophysical processes such as binary stalling or environmental coupling can modify the frequency distribution $P(f_r|M_{*}, {\rm z})$ \citep{sampson2015constraining,kelley2017massive,chen2017efficient}. The occupation fraction of GW sources within each frequency bin, expressed in terms of the spectral parameters $\alpha$ and $\lambda$, effectively encapsulates these effects and their evolution with redshift.

\begin{figure*}
    \centering
    \includegraphics[width=0.8\textwidth, height=0.35\textheight]{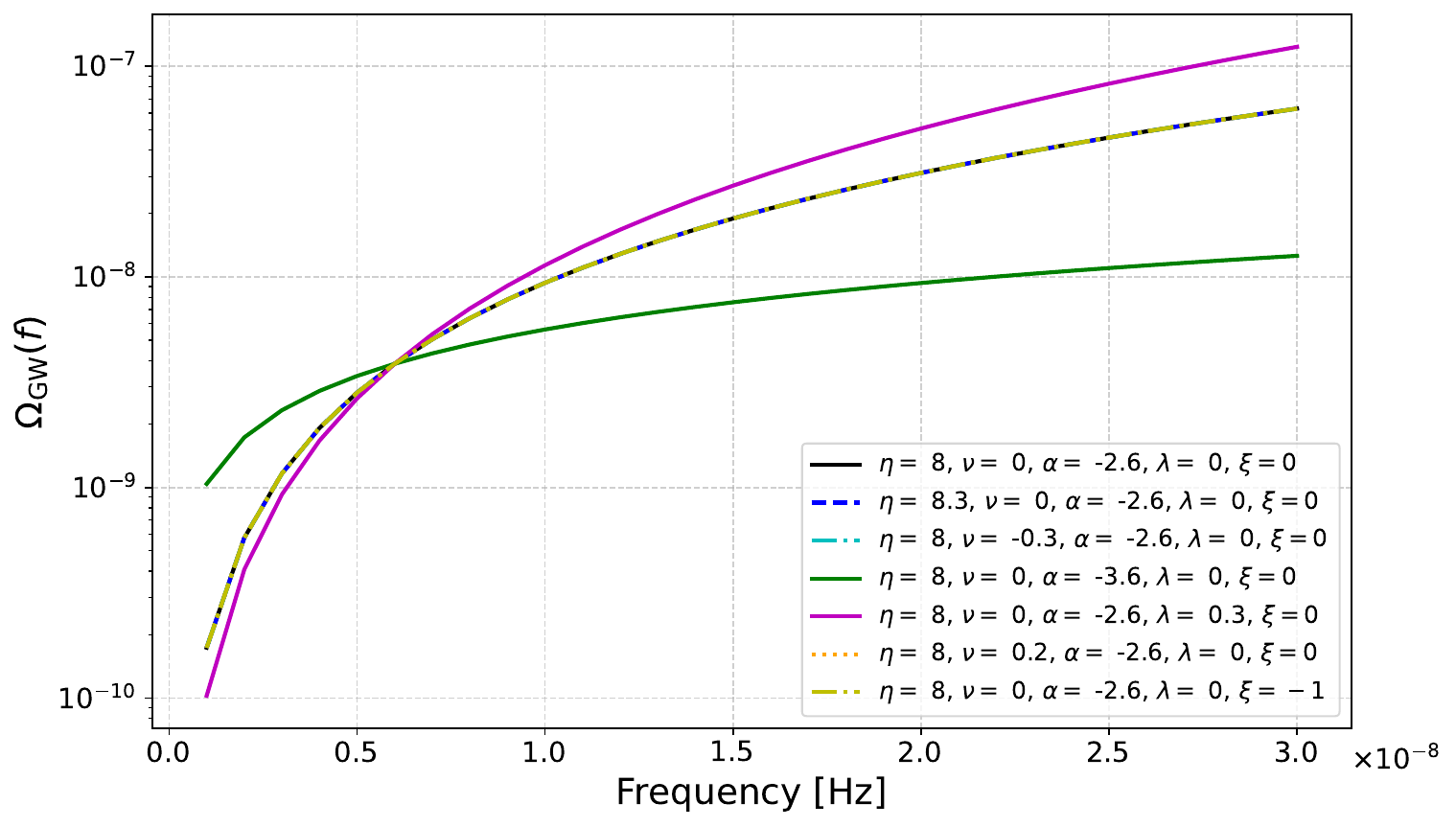}
    \caption{The SGWB energy density spectrum, $\Omega_{\rm GW}(f)$, as a function of frequency for different SMBHB population models. Various population parameters influence the spectrum differently: $\eta$, $\nu$, and $\xi$ primarily determine the overall amplitude, while $\alpha$ and $\lambda$ control the shape of the spectrum.}
    \label{fig:Omega}
\end{figure*}

In Fig.~\ref{fig:Omega}, we show the SGWB spectral density, $\Omega_{\rm GW}(f)$, for different values of $ \eta $, $ \nu $, $ \alpha $, $ \lambda $, and $\xi$. The stellar mass distribution of galaxies, $P(M_{*},{\rm z})$, is modeled using a redshift-dependent Schechter function following \citet{mcleod2021evolution}, which captures the redshift evolution of the galaxy stellar mass function. This evolution, in turn, impacts the mass distribution of SMBHBs. The redshift dependence of the binary formation rate is encoded via a factor of $(1+{\rm z})^{\xi}$. We have considered a fiducial case with $\xi=0$, which captures the same redshift evolution of the binaries and stellar mass. The normalization constant, $\mathcal{N}$, is chosen such that the spectrum attains a fixed value at a reference frequency, $f_{\rm ref}$. In this case, we set this value to the median value of the power law curve at $f = 6 \times 10^{-9}$ Hz fitted to the 15-year data release of NANOGrav \citep{agazie2023nanograv}.

The curves in Fig.~\ref{fig:Omega} highlight the sensitivity of $ \Omega_{\rm GW}(f) $ to variations in the parameters. Notably, the spectrum is highly insensitive to changes in $ \eta $, $\nu$, and $\xi$. This is because these parameters control the amplitude of the signal, which is fixed by the amplitude at a reference frequency ($f_{\rm ref} = 6 \times 10^{-9}$ Hz). This behavior introduces a degeneracy between $ \eta $, $\rho$, $ \nu $ and $\xi$, making it challenging to distinguish their individual contributions using $ \Omega_{\rm GW}(f) $ alone. In contrast, both $ \alpha $ and $ \lambda $ influence the shape of the spectrum. The parameter $\eta$ sets the minimum mass of the BHs. $\nu$ controls the redshift evolution of the mass of the BHs. A negative $\nu$ implies that at higher redshifts, galaxies of similar mass tend to host relatively lower-mass BHs, whereas a positive $\nu$ suggests that at higher redshifts, galaxies of similar mass tend to host comparatively more massive BHs. The parameters $\alpha$ and $\lambda$ define the frequency distribution of BHs, with $\lambda$ governing its redshift evolution. A positive $\lambda$ implies that at higher redshifts, there is a relative increase in the number of BHs emitting at higher frequencies, whereas a negative $\lambda$ indicates that at higher redshifts, there are relatively more binaries emitting at lower frequencies.

\begin{figure*}
    \centering
    \includegraphics[width=0.8\textwidth, height=0.35\textheight]{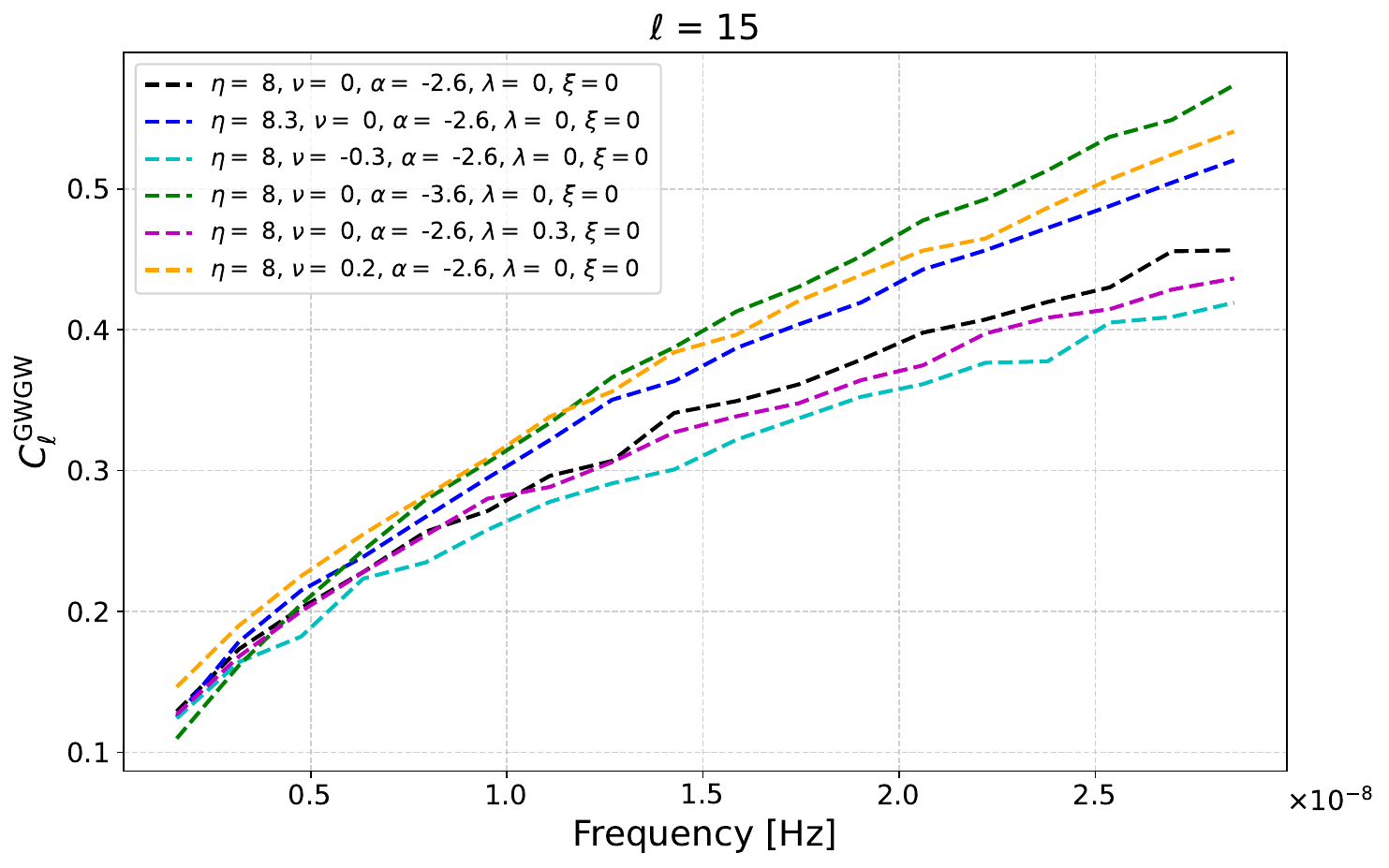} \
    \caption{The angular power spectrum of SGWB density, $C_{\ell}^{\rm GWGW}(f)$ at $\ell = 15$, as a function of frequency for different SMBHB population models. The $C_{\ell}^{\rm GWGW}(f)$ signal is expected to be shot noise-dominated, leading to a flat spectrum in spherical harmonic mode $\ell$.}

    \label{fig:CGWGW}
\end{figure*}

\section{Anisotropic GW Background Signal}\label{sec:Aniso}

This section describes the formalism of the anisotropic SGWB signal which includes the auto-angular power spectrum of the SGWB as well as its cross-correlation with the galaxy distribution. The cross-correlation is obtained between the spatial overdensity in galaxy number count ($\delta_{\rm g}(\hat{n})$) and the SGWB ($\delta_{\rm GW}(f,\hat{n})$). The spatial overdensity in the $\Omega_{\rm GW}(f,\hat{n})$ can be defined as \citep{sah2024imprints,sah2024discovering}

\begin{equation}
    \begin{aligned}
        \delta_{\rm GW}(f,\hat{n}) \equiv & ~ {\frac{   \Omega_{\rm GW}(f,\hat{n}) -  \overline{\Omega}_{\rm GW}(f)}{\overline{\Omega}_{\rm GW}(f)}}\\
         = &\int  d{\rm z} \frac{\int \prod\limits_{i}^{n} \frac{d\theta_i}{ d_{\ell}^{2} }  ~ \Big[\frac{dn}{d{\rm z}}~\overline{\kappa}(\Theta_{\rm n},f_{\rm r}|{\rm z}) \Big]~ \frac{ dE_{\rm gw}(f_{{\rm r}},\Theta_{\rm n})}{dt_{\rm r}}}{\int  d{\rm z} ~ \int \prod\limits_{i}^{n} \frac{d\theta_i}{ d_{\ell}^{2} }  ~\Big[\frac{dn}{d{\rm z}}~\overline{\kappa}(\Theta_{\rm n},f_{\rm r}|{\rm z}) \Big] ~ \frac{ dE_{\rm gw}(f_{{\rm r}},\Theta_{\rm n})}{dt_{\rm r}} } \Big[\frac{\kappa(\Theta_{\rm n},f_{\rm r},\hat{n}|{\rm z})\frac{dn}{d\rm{V}}({\rm z}, \hat{n})}{\overline{\kappa}(\Theta_{\rm n},f_{\rm r}|{\rm z}) \overline{\frac{dn}{d\rm{V}}}({\rm z})} - 1\Big]\\
        = & \int ~d{\rm z}~ 
     ~  \phi_{\rm GW}(f,{\rm z}) \times [\delta_{\rm m}({\rm z},\hat{n}) ~ b_{\rm GW}({\rm z})],
    \label{SGWB_Fluc}
    \end{aligned}
\end{equation}
where $\delta_{\rm m}({\rm z},\hat{n})$ is the matter density fluctuation, $b_{\rm GW}({\rm z})$ is the GW bias, quantifying how GW sources trace trace the underlying matter distribution in the Universe, $\frac{dn}{d{\rm z}}$ represents the redshift distribution of the galaxy, and $\phi_{\rm GW}(f,{\rm z})$ is the GW window function given by
\begin{equation}
    \phi_{\rm GW}(f,{\rm z})  = \frac{\int\prod\limits_{i}^{n} \frac{d\theta_i}{ d_{\ell}^{2} }  \Big[\frac{dn}{d{\rm z}} ~\overline{\kappa}( \Theta_{\rm n}, f_r|{\rm z} ) \Big] \frac{ dE_{\rm gw}(f_r,\Theta_{\rm n})}{dt_{\rm r}}}{\int d{\rm z} \int  \prod\limits_{i}^{n} \frac{d\theta_i}{ d_{\ell}^{2} }  \Big[\frac{dn}{d{\rm z}}~\overline{\kappa}(\Theta_{\rm n}, f_r|{\rm z} )\Big] \frac{ dE_{\rm gw}(f_r,\Theta_{\rm n})}{{ dt_{\rm r}}}},
    \label{phiGW}
\end{equation}

The fluctuations in the SGWB ($\delta_{\rm GW}(\hat{n})$) and galaxy density ($\delta_{\rm g}(\hat{n})$) can be expressed in the spherical harmonic basis as
\begin{align}
    \delta_{\rm g,GW}(\hat{n}) =& \sum\limits_{\ell} \sum\limits_{m = -\ell}^{\rm \ell}  c_{\ell m}^{\rm g,GW} Y_{\ell m}(\hat{n}).
\end{align}
The angular power spectrum of the SGWB is expected to be dominated by shot noise \citep{sah2024imprints}. This behavior arises due to the discrete nature of the GW sources, where the finite number of contributing sources introduces stochastic fluctuations that overshadow any intrinsic clustering signal.
The auto-angular power spectrum for the shot noise-dominated signal can be written as
\begin{equation}
    C_{\ell}^{\rm GWGW}(f) \sim \big<\delta^{2}_{\rm GW}(f,\hat{n})\big>,
    \label{CGWGW}
\end{equation}
and the uncertainty in the measurement of the $C_{\ell}^{\rm GWGW}(f)$ signal, $\Sigma_{\rm GWGW}^{2}(f,\ell)$, for a Gaussian signal is given by

\begin{equation}
    \begin{aligned}
        \Sigma_{\rm GWGW}^{2}(f,\ell) =  &\frac{1}{  f_{\rm sky} (2 \ell+1)}  \bigg[ \Big(C_{\ell}^{\rm GWGW}(f) + N_{\ell}(f)\Big)^{2}\bigg], 
    \end{aligned}
    \label{Var1}
\end{equation}

where $N_{\ell}$ is the uncertainties in the measurement of $C_{\ell}^{\rm GWGW}(f)$ due to timing residual noise, $f_{\rm sky}$ denotes the fraction of the sky covered by the galaxy catalog. We adopt a multipole bin width of $\Delta\ell = 1$, while the frequency bin width is determined by the total observation time of the PTA, $T_{\rm obs}$, such that $\Delta f = 1/T_{\rm obs}$. For LSST, $f_{\rm sky} \approx 0.42$, and we assume the same value throughout the rest of this paper.

In Fig. \ref{fig:CGWGW}, we show the $C_{\ell}^{\rm GWGW}(f)$ as a function of frequency. The overall shape of the spectrum is relatively insensitive to variations in the population model. The case with a large value of $\eta$ requires a relatively smaller number of sources to generate the same power spectrum. Hence, in this scenario, the anisotropy is expected to be larger, as evident from the figure. Similarly, the case where heavier BHs reside at higher redshifts (positive $\nu$) results in a higher amplitude compared to the scenario with lighter BHs at higher redshifts, due to the same underlying reason as the previous case. In cases with more negative $\alpha$ and small $\lambda$, more BHs are emitting at low frequencies. Hence, the curve for the case of $\alpha = -3.6$ is above the $\alpha = -2.6$ at high frequencies and below $\alpha = -2.6$ at low frequencies.

The angular cross-correlation between the SGWB map and the galaxy map can be defined as \citep{sah2024discovering}
\begin{equation}
    \begin{aligned}
       C_{\ell}^{\rm gGW}(f) \equiv\, &
       \langle c_{\ell m}^{\rm GW}(f)~  c_{\ell 
        m}^{\rm g} \rangle,\\
         =& \frac{2}{\pi}  \int d{\rm z}_1 b_{g}({\rm z}_1)\phi_{\rm g}({\rm z}_1) D({\rm z}_1) \int dz_2 b_{\rm GW}({\rm z}_2) \phi_{\rm GW}(f,{\rm z}_2) \\ & \times ~ D({\rm z}_2)
          \int k^2 dk ~  j_{\ell}(kr({\rm z}_1)) j_{\ell}(k r({\rm z}_2))  ~ P(k),
    \end{aligned}
    \label{ClgGW_1}
\end{equation}
where $b_{g}({\rm z})$ is the galaxy bias with respect to the underlying matter distribution. For this analysis, we assume $b_{\rm GW}({\rm z}) = b_{g}({\rm z})$, adopting the galaxy bias inferred from \citet{pandey2025cosmology}. $j_{\ell}(kr)$ are spherical Bessel functions of order $\ell$, and $P(k)$ is the matter power spectrum at redshift z=0. The term $D({\rm z})$ represents the growth factor of density perturbations over time, $\phi_{\rm g}({\rm z}) \equiv \frac{\frac{dn}{d{\rm z}}}{\int \frac{dn}{d{\rm z}} ~ d{\rm z}}$  is the galaxy window function. The cross-correlation can also be obtained between the individual tomographic bin and the frequency bin of the SGWB density.
\begin{equation}
   C_{\ell}^{\rm gGW}(f,{\rm z}) \equiv\, 
   \langle c_{\ell m}^{\rm GW}(f)~  c_{\ell 
    m}^{\rm g}({\rm z}) \rangle,
    \label{ClgGW_z}
\end{equation}
where  $C_{\ell}^{\rm gGW}(f,{\rm z})$ is the cross power spectrum between SGWB density measured in a frequency bin ($\Delta f$) around the frequency $f$ and the galaxy in a redshift bin ($\Delta {\rm z}$) around ${\rm z}$, $c_{\ell m}^{\rm g}({\rm z})$ is the spherical harmonic component of the galaxy number density in a redshift bin around z. The uncertainty in the measurement of the $C_{\ell}^{\rm gGW}(f,{\rm z})$ signal, $\Sigma_{\rm gGW}^{2}(f,\ell,{\rm z})$, in the Gaussian limit is given by
\begin{equation}
    \begin{aligned}
        \Sigma_{\rm gGW}^{2}(f,\ell,{\rm z}) =  &\frac{1}{f_{\rm sky}   (2 \ell+1)}  \bigg[(C_{\ell}^{\rm gGW}(f,{\rm z}))^2 \\ &  + (C_{\ell}^{\rm gg}({\rm z}) + \frac{1}{n_{\rm g}})  (C_{\ell}^{\rm GWGW}(f) + N_{\ell}(f))\bigg], 
    \end{aligned}
    \label{Var2}
\end{equation}
where $C_{\ell}^{\rm GWGW}$ and $C_{\ell}^{\rm gg}$ represent the angular power spectra of SGWB and galaxy, respectively. The term $n_{\rm g}$ corresponds to the mean galaxy number per steradian. 

\begin{figure*}
    \centering
    \includegraphics[width=0.8\textwidth, height=0.35\textheight]{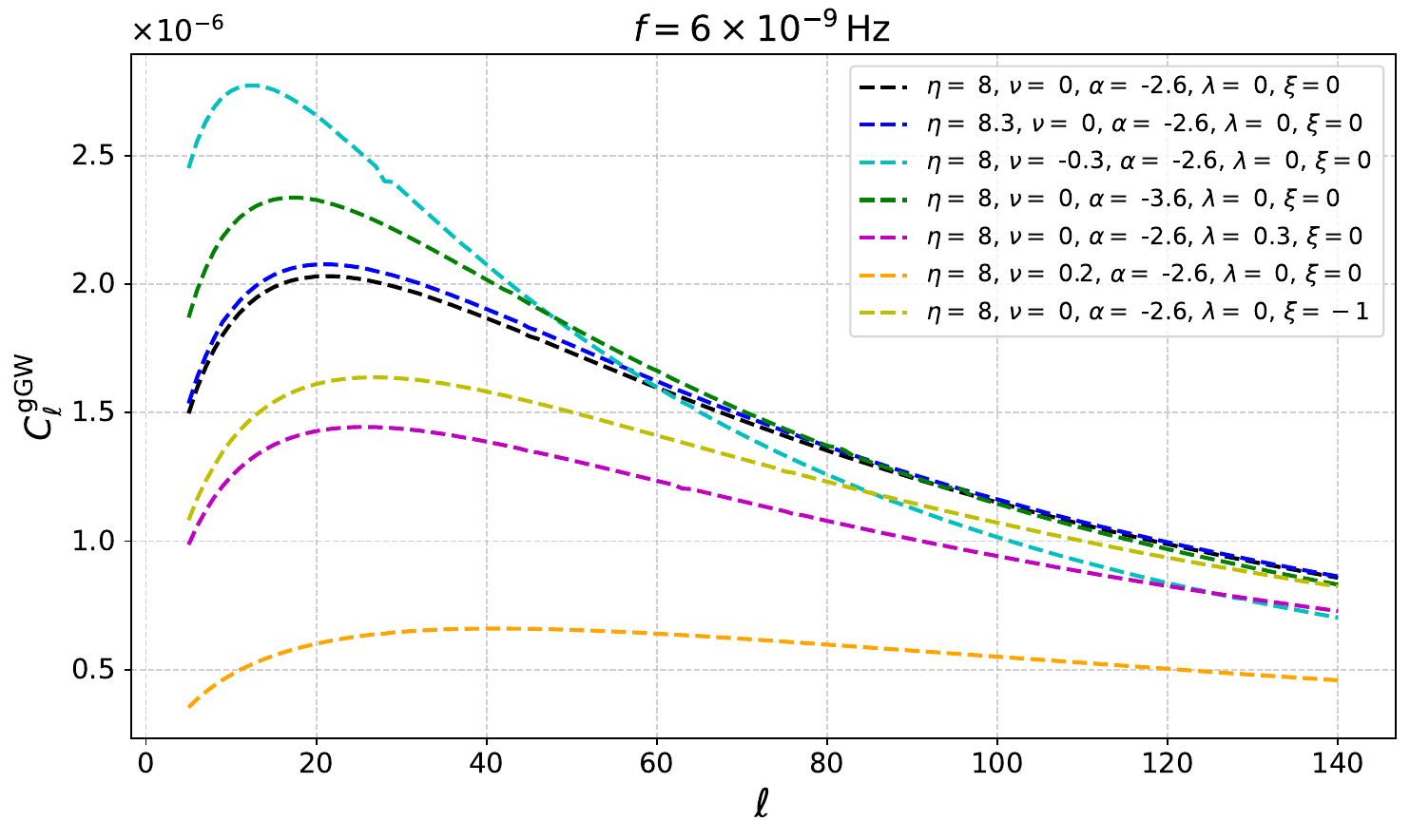} 
  
    \caption{The cross-correlation power spectrum between SGWB density and the galaxy distribution, $C_{\ell}^{\rm gGW}(f)$, as a function of spherical harmonic mode $\ell$ at a GW frequency of $6 \times 10^{9}$ Hz, for different population models. The shape and structure of the signal are primarily influenced by the population parameters $\nu$, $\alpha$, $\lambda$, and $\xi$ while $\eta$, which controls the overall amplitude of $\Omega_{\rm GW}(f)$, has little to no effect on $C_{\ell}^{\rm gGW}$.}
    \label{fig:ClgGW}
\end{figure*}

\begin{figure*}
    \centering
    \includegraphics[width=0.8\textwidth, height=0.35\textheight]{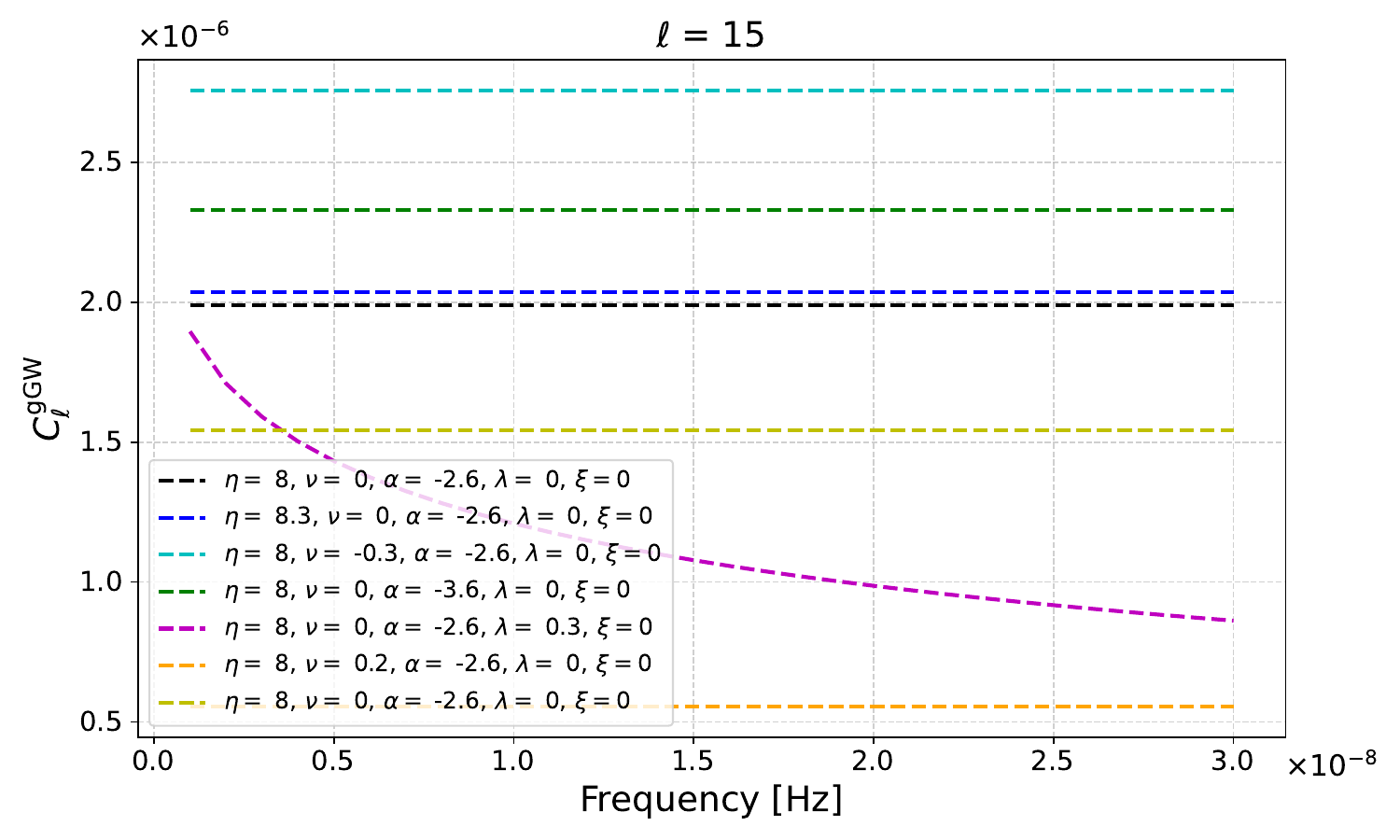} 
  
    \caption{The cross-correlation power spectrum between SGWB density and the galaxy distribution, $C_{\ell}^{\rm gGW}$ at $\ell = 15$, as a function of GW frequency for different population models. The curves are flat for all cases except when $\lambda \neq 0$. This is because the window function $\phi_{\rm GW}(f,{\rm z})$ is independent of frequency in cases where the frequency distribution does not depend on redshift.}
    \label{fig:ClgGW_f}
\end{figure*}

In Fig. \ref{fig:ClgGW}, we illustrate the cross-correlation signal, $C_{\ell}^{\rm gGW}(f)$, between SGWB (at $f = 6 \times 10^{-9}$ Hz) and galaxy distribution (redshift integrated galaxy map) of LSST-like surveys, for different values of $\eta$, $\nu$, $\alpha$, $\lambda$, and $\xi$. The $C_{\ell}^{\rm gGW}(f)$ spectrum depends on the relative contribution of the SGWB and the galaxy density from different redshifts. The $C_{\ell}^{\rm gGW}(f)$ is largely unaffected by variations in the $M_{*}-M_{\rm BH}$ relation (i.e., to changes in $\eta$). However, it exhibits significant changes when redshift evolution is incorporated into the $ M_{*}-M_{\rm BH} $ ($\nu\neq0$). For instance, with $\nu = -0.3$, the SMBHs are relatively less massive at high redshifts compared to the low redshifts. This means the relative contribution of the SGWB signal is enhanced at lower redshifts.  Similarly, in Fig. \ref{fig:ClgGW_f}, we show the $C_{\ell}^{\rm gGW}$ as a function of frequency for $\ell = 15$. The curves are flat for all cases except for the case where $\lambda \neq 0$. This is because the window function ($\phi_{\rm GW}(f,{\rm z})$) is independent of the frequency for all cases where the frequency distribution is independent of the redshift.

The cross-angular power spectrum can be interpreted as the projection of the matter power spectrum onto the window function. This implies that if both the window function and the power in the matter power spectrum corresponding to a given $\ell$-mode exhibit a similar trend with redshift, the power in that $\ell$-mode will be enhanced. The Fourier modes (k-modes) associated with a given angular multipole $\ell$ decrease with redshift. Under the Limber approximation, they are given by, $k = \frac{\ell + 1/2}{d_m({\rm z})}$, where $d_m({\rm z})$ represents the comoving distance. At low $\ell$ values, this decrease corresponds to a reduction in the clustering power of matter as redshift increases.  For negative values of $\nu$, where the window function declines more steeply with redshift, the projection factor becomes large. Consequently, the angular power spectrum $C_{\ell}^{\rm gGW}(f)$ is expected to be larger at low $\ell$. However, at higher $\ell$ values, the increase in redshift corresponds to an increase in the clustering power of matter, thereby reducing the projection factor. As a result, for these higher $\ell$ values, $C_{\ell}^{\rm gGW}(f)$ decreases as $\nu$ decreases.

The steepness of the frequency distribution, governed by $ \alpha $ and $\lambda$, also affects $ C_{\ell}^{\rm gGW}(f) $. A more negative $ \alpha $ implies a steeper frequency distribution, causing more sources emitting at lower frequencies to be redshifted out of the observational band, decreasing the value of the window function at higher redshifts. This behavior impacts the $ C_{\ell}^{\rm gGW}(f) $ spectrum in a manner similar to $ \nu $, as both parameters influence the relative redshift contributions.

\begin{figure*}
    \centering
    \includegraphics[width=0.9\textwidth, height=0.4\textheight]{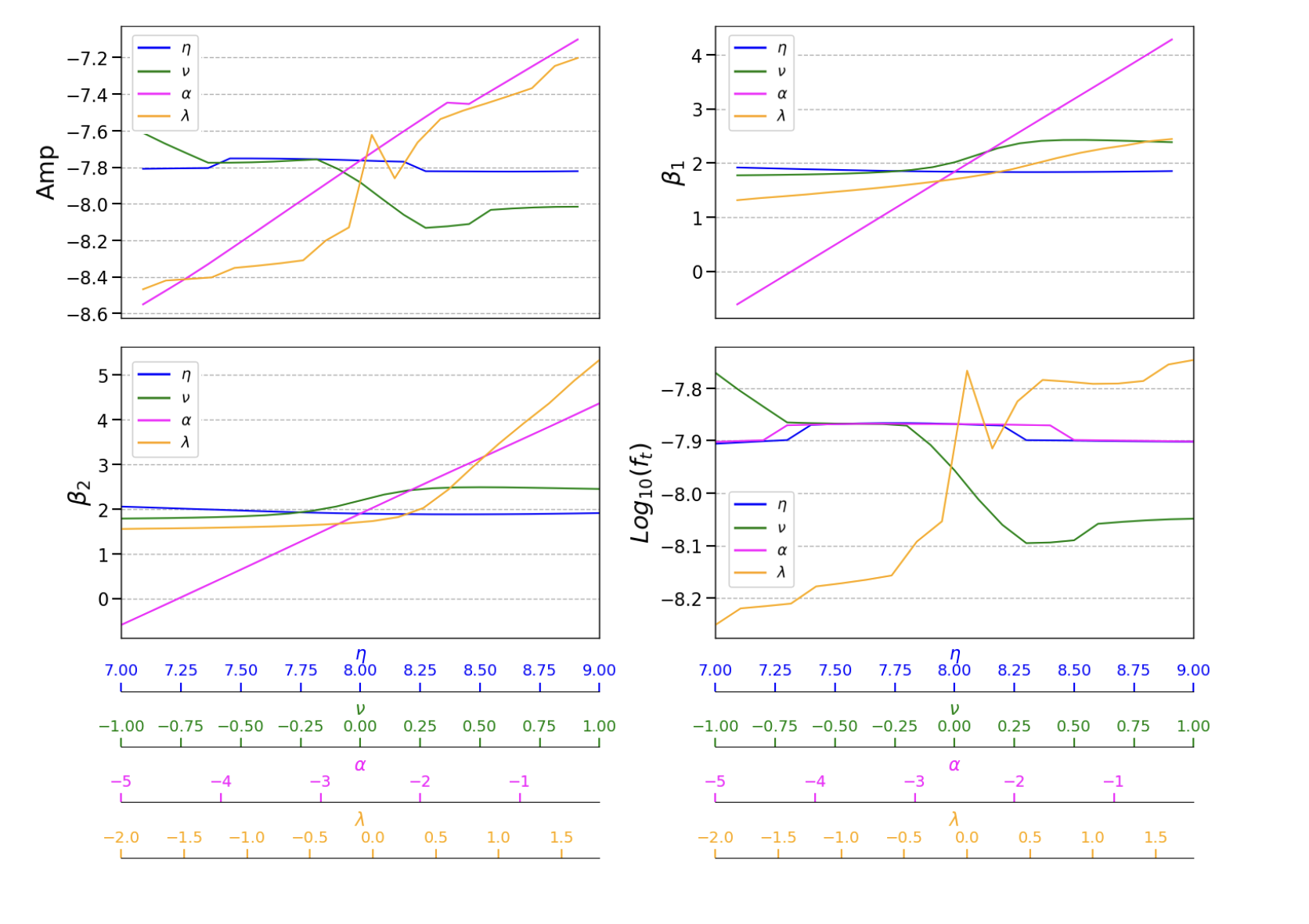} 
  
    \caption{Mapping between the SGWB power spectrum ($\Omega_{\rm GW}(f)$) features and physical parameter, $\eta$, $\nu$, $\alpha$ and $\lambda$. The $\log_{10}(\Omega_{\rm GW}(f))$ is fitted to linear curve with an intercept $Amp$ and slopes $\beta_1$ and $\beta_2$ with a parameters $f_t$ denoting the transition frequency of the curve from slope $\beta_1$ to $\beta_2$.}
    \label{Par_Phy_Omega}
\end{figure*}

\begin{figure*}
    \centering
    \includegraphics[width=0.9\textwidth, height=0.4\textheight]{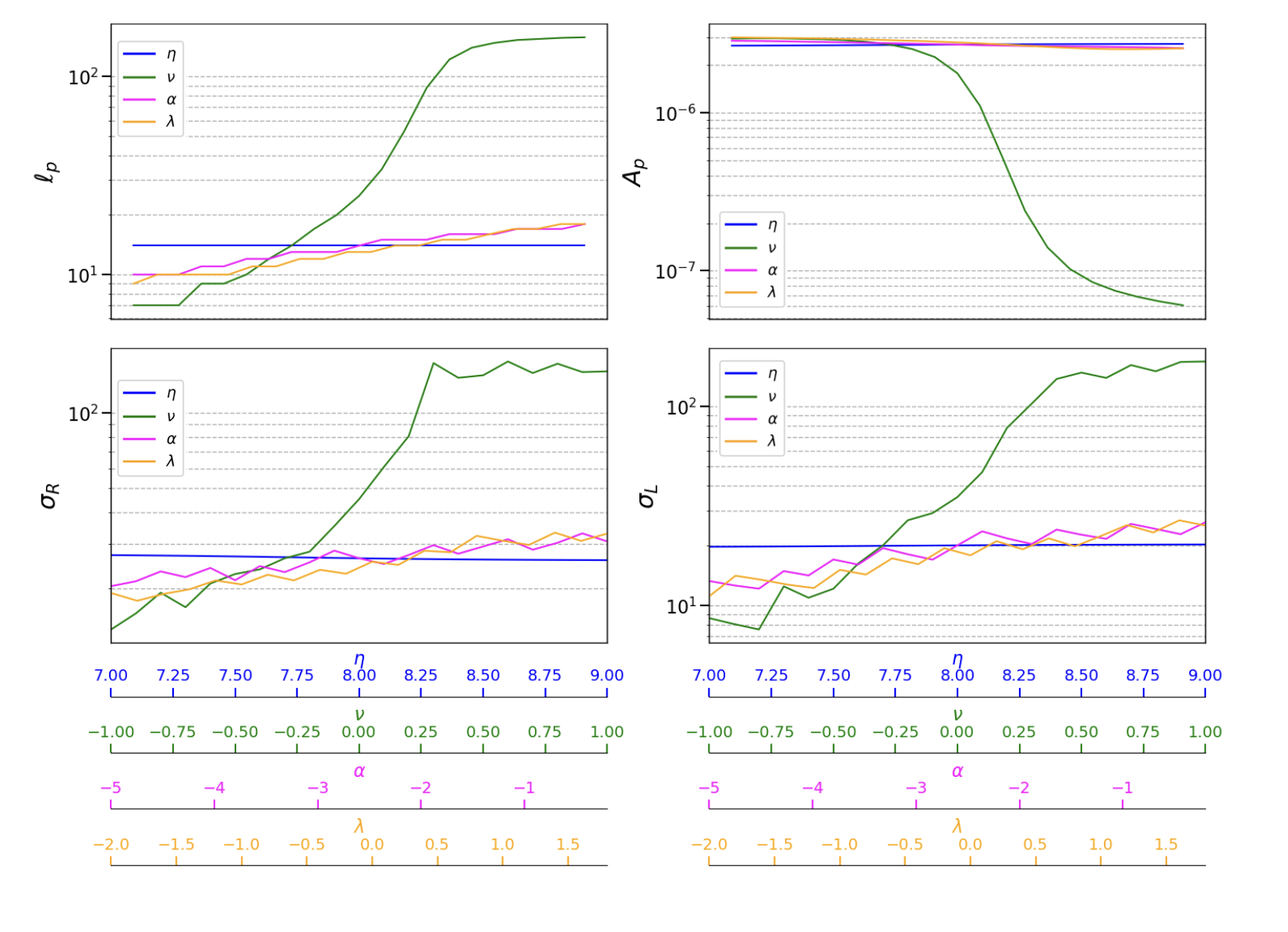}
  
    \caption{Mapping between the cross-angular power spectrum ($C_{\ell}^{\rm gGW}$), at frequency  $6 \times 10^{-9}$ Hz. The $C_{\ell}^{\rm gGW}$ is fitted 
    with one-sided Gaussian to the left and right sides of the peak of $C_{\ell}^{\rm gGW}$. The $\ell_p$ denotes the $\ell$ value at which $C_{\ell}^{\rm gGW}$ peaks, $A_p$ is the height of the peak, and $\sigma_L$ and $\sigma_R$ are the standard deviation of the left and right Gaussian, respectively.}
    \label{Par_Phy1}
\end{figure*}

\section{Mapping \lowercase{{\large between}} observational quantities \lowercase{{\large and}} physical parameters}\label{sec:Map}
The nHz SGWB encodes valuable information about the population and evolution of SMBHBs. The observable quantities, such as the SGWB spectral density $\Omega_{\rm GW}(f)$, the angular power spectrum $C_\ell^{\rm GWGW}(f)$, and the cross-angular power spectrum $C_\ell^{\rm gGW}(f)$, exhibit variations in amplitude and shape across spatial ($\ell$) and spectral ($f$) frequencies. These variations are governed by underlying physical parameters, including the SMBHB mass distribution ($\eta$, $\rho$, $\nu$), and the parameters controlling the frequency distribution of the GW sources ($\alpha$, $\lambda$). Understanding the dependence of these observables on the physical parameters is crucial for determining which aspects of the SMBHB population can be constrained from PTA observations. In this section, we examine how these parameters affect the observed signals and discuss their implications for constraining the SMBHB population. We particularly focus on $\Omega_{\mathrm{GW}}(f)$ and $C_{\ell}^{\mathrm{GW}}(f)$ in this section, as these two quantities mainly contribute to constraining the physical parameters of interest in this analysis.

\begin{enumerate}
    \item \textbf{Impact of Physical Parameters on $\Omega_{\rm GW}(f)$:} The slope of the $\Omega_{\rm GW}(f)$ curve may deviate from a simple power law due to the redshift evolution of the frequency distribution. To account for this, we fit a power law to $\Omega_{\rm GW}$, parameterized by an amplitude $Amp$, a transition frequency $f_t$, and two spectral indices $\beta_1$ and $\beta_2$, such that
\begin{equation}
     \log_{10}\big(\Omega_{\rm GW}(f)\big) \equiv \bigg\{
    \begin{array}{cl}
    &Amp  + \beta_1 \big(\log_{10}(f/f_t)\big), \quad  f \leq f_{t},\\
    & Amp  + \beta_2 \big(\log_{10}(f/f_t)\big), \quad f > f_{t}
    \end{array}
    \label{eq:broken_powerlaw}
\end{equation}
These parameters are the observable quantities that can demonstrate the observed features in the SGWB power spectra. In Fig. \ref{Par_Phy_Omega}, we show how the observable quantities, $Amp$, $\beta_1$, $\beta_2$ and $f_{t}$, are affected by the physical parameters like $\eta$, $\nu$, $\alpha$, and $\lambda$. We adopt the fiducial values $\eta = 8$, $\rho = 1$, $\nu = -0.3$, $\xi = 0$, $\alpha = -2.6$, and $\lambda = 0.3$, and vary the physical parameters one at a time to study their impact on the observables. The effects of different physical parameters on these features are summarized below.

\begin{itemize}
    \item \textit{Overall Amplitude ($Amp$) and Transition Frequency ($f_t$):} Since we have normalized $\Omega_{\rm GW}(f)$ such that it remains fixed at a reference frequency ($f_{\rm ref} = 6 \times 10^{-9}$ Hz), $Amp$ and $f_t$ are unaffected by variations in $\eta$. However, it is strongly influenced by the frequency distribution parameters, $\alpha$, and $\lambda$, both of which lead to an increase in $Amp$ as their values increase.  A positive value of $\nu$ increases the contribution from high-redshift sources, where the frequency distribution is shallower. As a result, the spectrum becomes steeper, shifting $f_t$ to a lower value and consequently, decreasing $Amp$.

    \item \textit{Spectral Slopes ($\beta_1, \beta_2$)}: The positive change in both $\alpha$ and $\lambda$ results in the steeper $\Omega_{\rm GW}(f)$, as a result both $\beta_1$ and $\beta_2$ increase with the increase in $\alpha$ and $\lambda$. The parameters $\nu$ do not affect the spectral slope of the $\Omega_{\rm GW}(f)$ significantly.

\end{itemize}

These findings indicate that the spectral shape of $\Omega_{\rm GW}(f)$ is mostly insensitive to the parameter $\eta$ and there exists significant degeneracy between $\eta$ and $\nu$. However, $\Omega_{\rm GW}(f)$ shows that spectral features are strongly affected by changes in $\alpha$ and $\lambda$. This implies that $\Omega_{\rm GW}(f)$ can effectively constrain the frequency evolution of SMBHB.

\item \textbf{Impact of Physical Parameters on $C_\ell^{\rm gGW}(f)$:} Similarly, we fit the cross-correlation spectrum $C_\ell^{\rm gGW}(f)$ at a fixed frequency ($6 \times 10^{-9}$ Hz) using a Gaussian model.

\begin{equation}
    C_{\ell}^{\rm gGW} \equiv A_{p} ~ \exp \left(\frac{-(\ell - \ell_{p})^2}{2 \sigma_{L,R}^2}\right),
\end{equation}
where, $\ell_p$ is the multipole where $C_\ell^{\rm gGW}$ peaks, $A_p$ is the peak amplitude, $\sigma_L$, and $\sigma_R$ are the standard deviations of the one-sided Gaussian fitted to the left and right of the peak, respectively. For fitting, we select five consecutive points around the peak.

In Fig. \ref{Par_Phy1}, we demonstrate how the structure of the $C_{\ell}^{\rm gGW}$ is impacted by the physical parameters. We adopt fiducial values of $\eta = 8$, $\rho = 1$, $\nu = -0.3$, $\alpha = -2.6$, $\lambda = 0.3$, and $\xi = 0$ and vary one parameter at a time while keeping the others fixed. The effects of different physical parameters on these features are summarized below. 

\begin{itemize}
    \item \textit{Peak Location ($\ell_p$)}: As expected from earlier discussions, the parameter $\eta$ has a negligible effect on the cross-correlation features. In contrast, $\nu$ has a significant impact, shifting $\ell_p$ to higher values as $\nu$ increases. This shift occurs because a higher $\nu$ reduces the relative contribution from lower redshifts and increases the power in higher $\ell$ modes, as discussed in Sec. \ref{sec:Aniso}.  The impact of $\alpha$ and $\lambda$ on $\ell_{p}$ is relatively small. However, $\ell_p$ shifts toward higher values with an increase in both parameters. This occurs because the frequency distribution becomes shallower as these parameters increase, resulting in fewer sources being redshifted out of the observation band. Consequently, the window function attains relatively larger values at higher redshifts compared to a steeper frequency distribution.  By the same reasoning as in Sec.~\ref{sec:Aniso}, the angular power spectrum $C_{\ell}^{\rm gGW}$ at lower $\ell$ values decreases with increasing $\alpha$ and $\lambda$. This implies that $\ell_p$ shifts to a higher value with the increase in these parameters.

    \item \textit{Peak Amplitude ($A_p$)}: The peak amplitude $A_p$ decreases as $\nu$ increases, as the overall amplitude of the SGWB diminishes with an increasing contribution from higher redshifts, as discussed in Sec. \ref{sec:Aniso}. The effect of $\alpha$ and $\lambda$ on $A_p$ is relatively weak.

    \item \textit{Peak Width ($\sigma_L$, $\sigma_R$)}: The parameter $\nu$ has a strong impact on both $\sigma_L$ and $\sigma_R$, which increase as $\nu$ increases. This trend arises because a larger $\nu$ enhances the contribution from high-redshift sources in the window function, leading to a shallower decline of $C_{\ell}^{\rm gGW}$ with $\ell$. Similarly, $\sigma_L$ and $\sigma_R$ also grow with increasing $\alpha$ and $\lambda$, since a shallower frequency distribution yields a greater weight from high-redshift sources, again making the falloff of $C_{\ell}^{\rm gGW}$ with $\ell$ less steep.  
\end{itemize}

\end{enumerate}

These results demonstrate that $\nu$ is the key parameter influencing the cross-correlation power spectrum, highlighting the crucial role of $C_{\ell}^{\rm gGW}$ in breaking the degeneracy present in $\Omega_{\rm GW}(f)$ among $\eta$, $\rho$, and $\nu$. While both $\Omega_{\rm GW}(f)$ and $C_{\ell}^{\rm gGW}$ are sensitive to $\alpha$ and $\lambda$, these parameters exhibit some degeneracy in both signals individually. However, their combination significantly improves constraints, providing more precise measurements of these quantities.

\section{Cross-Correlation Analysis \lowercase{{\large in}} Pulsar Timing Arrays}\label{sec:Cross}
PTAs are uniquely capable of detecting the SGWB by correlating the timing residuals of the pulse arrival time of multiple millisecond pulsars. These residuals are influenced by passing GWs, which induce correlated deviations in the pulse arrival times. The cross-correlation of these timing residuals provides a powerful tool to probe GW signals and differentiate them from uncorrelated noise sources.

The timing residual in the timing signal of a pulsar located in the direction $\hat{p}$ due to the GW propagating in the direction $\hat{n}$ is given by \citep{anholm2009optimal,chamberlin2015time,hobbs2017gravitational,burke2019astrophysics,maiorano2021principles}
\begin{equation}
r(t,\hat{n}) = \int_0^t dt' \frac{1}{2} \frac{\hat{p}_i \hat{p}_j}{1 + \hat{n} \cdot \hat{p}} \Big(h_{ij}(t', \hat{n}) - h_{ij}(t' - t_p, \hat{n})\Big),
\end{equation}
where $h_{ij}$ is the metric perturbation, $t_p$ is the light travel time between the pulsar and the earth. The metric perturbation from an SGWB can be expanded as:
\begin{equation}
    h_{ij}(t, \vec{\rm x}) = \sum_A  \int d\hat{n} ~ \int_{-\infty}^\infty df ~h_A(f, \hat{n})~ e^{i 2 \pi f (t - \hat{n} \cdot \mathbf{\vec{x}})} ~ e^A_{ij}(\hat{n}),
\end{equation}
where \(A = +, \times\) denotes the polarization modes and \(e^A_{ij}(\hat{\Omega})\) are the polarization tensors.

This expansion allows us to express the timing residuals in the frequency domain 
\begin{equation}
    \tilde{r}(f, \hat{n}) = \frac{1}{2\pi i f} \left( 1 - e^{-2\pi i f L (1 + \hat{n} \cdot \hat{p})} \right) 
    \sum_A h_A(f, \hat{n}) ~  \mathcal{F}^{A}(\hat{p},\hat{n}) ,
    \label{res_freq}
\end{equation}

The expectation value of the metric perturbation, $\langle h^*_A(f, \hat{n}) h_{A}(f', \hat{n}') \rangle$ can be expressed as a function of strain spectral density, $S_{h}(f,\hat{n})$

\begin{equation}
    \langle h^*_A(f, \hat{n}) h_{A}(f', \hat{n}') \rangle = \delta(f - f')  ~ \delta^2(\hat{n} - \hat{n}')~ \frac{1}{16 \pi}  ~S_{h}(f,\hat{n}). 
    \label{strain_cross}
\end{equation}
Using Eq. \eqref{res_freq} and Eq. \eqref{strain_cross}, we can express the cross-correlation of the Fourier transform of timing residual between two pulsars, denoted by I and J as
\begin{equation}
    \begin{aligned}
        \langle \tilde{r}_I(f) \tilde{r}_J(f') \rangle & = \frac{1}{24 \pi^2} \delta(f - f') f^{-2} \int S_{h}(f,\hat{n}) ~ \Gamma_{IJ}(\hat{n}) ~ d\hat{n}, \\
        & = \frac{1}{24 \pi^2} \delta(f - f') f^{-2} S_{h}(f) \int P(f,\hat{n}) ~ \Gamma_{IJ}(\hat{n}) ~ d\hat{n}, 
    \end{aligned}
\end{equation}
where $S_{h}(f,\hat{n}) \equiv S_{h}(f) P(f,\hat{n})$, and $S_{h}(f) \equiv \frac{1}{4 \pi}\int\limits_{S^2} d\hat{n} ~ S_{h}(f,\hat{n})$. The spectral density, $S_{h}(f)$ can be written in terms of the characteristic strain ($h_c(f)$) and $\Omega_{\rm GW}(f)$

\begin{equation}
    h^2_{c}(f) = f ~ S_{h}(f), 
\end{equation}

\begin{equation}
    \Omega_{\rm GW}(f) = \frac{2 \pi^2}{3 H_0^2} ~ f^3 ~ S_{h}(f). 
\end{equation}

In addition to the GW signal, timing residuals are affected by uncorrelated noise contributions. The timing residual noise power spectrum can be modeled as 
\begin{equation}
    P(f) = P_{\rm w}(f) + P_{\rm r}(f),
\end{equation}
where $P_{\rm w}(f) = 2~ \Delta t \sigma^{2}_{\rm w} $ is the white noise. Here, $\Delta t$ denotes the cadence of the pulsar time-of-arrival (TOA) measurements, and $\sigma_{\rm w}$ represents the uncertainty in the TOA measurements. $P_{\rm r}(f)$ is the red-noise power spectrum which is composed of two components: $P_{\rm I}(f)$, the intrinsic red noise, and $P_{\rm gw}(f)$, the GW-induced noise 
\begin{equation}
    P_{\rm r}(f) = P_{\rm I}(f) + P_{\rm gw}(f).
\end{equation}
The intrinsic red noise, $P_{\rm I}(f)$, is modeled as
\begin{equation}
    P_{\rm I}(f) = \frac{A^{2}_{\rm rd}}{12\pi^2} ~ \Big(\frac{f}{f_{\rm 1yr}}\Big)^{\gamma} f^{-3}_{\rm 1yr},
\end{equation}
where $A_{\rm rd}$ is the amplitude of the intrinsic red noise, \(f_{\rm 1yr}\) is the reference frequency corresponding to one year, $f_{\rm 1yr} = 1/\text{yr}$, and $\gamma$ is the spectral index of the red noise. The GW-induced noise is expressed as:
\begin{equation}
    P_{\rm gw}(f) = \frac{h_c^2}{12 \pi^2} f^{-3},
\end{equation}
where $h_c$ is the characteristic strain of the GW background. This formulation provides a comprehensive decomposition of the total power spectrum into white noise and red noise components, enabling the modeling of both instrumental effects and astrophysical contributions to the timing residuals.

In addition to the white noise and red noise components discussed above, other potential sources of noise can impact pulsar timing residuals. These include lensing noise, which arises due to the gravitational lensing of pulsar signals by intervening massive structures, and pulsar glitch noise, which results from sudden changes in the rotational frequency of pulsars \citep{lower2020utmost,goncharov2021identifying,reardon2023gravitational,falxa2024modeling,di2024systematic}. Lensing noise can introduce frequency-dependent distortions in the timing residuals, while pulsar glitches are typically analyzed on an individual pulsar basis, as their effects are highly specific to each pulsar's internal dynamics. In this analysis, these noise sources are not included, as their contributions are expected to be subdominant compared to the dominant white and red noise and GW-induced noise at the frequencies of interest. However, future work will incorporate these additional noise sources to provide a more comprehensive characterization of the timing residuals and their impact on SGWB detection.

\begin{figure*}
    \centering
    \includegraphics[width=0.7\textwidth, height=0.5\textheight]{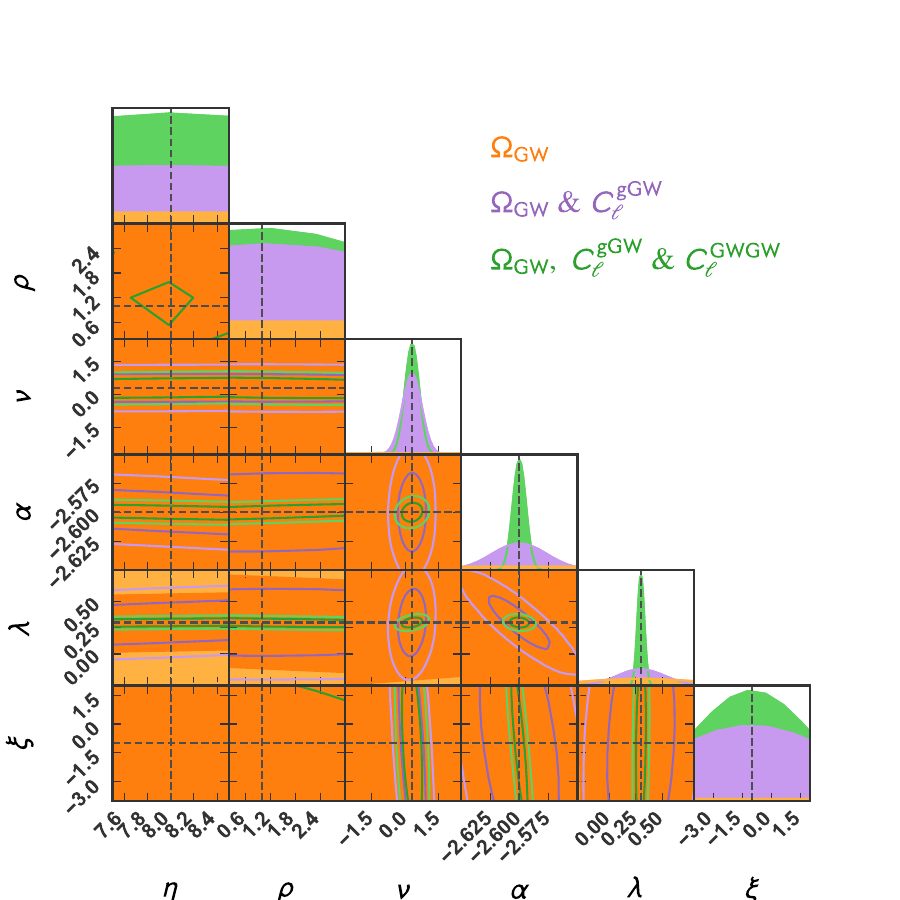} 
    \caption{Corner plot showing the constraints on the parameters $\eta$, $\rho$, $\nu$, $\alpha$, $\lambda$, and $\xi$ for three scenarios: (i) using only $\Omega_{\rm GW}(f)$, (ii) including $C_{\ell}^{\rm gGW}(f)$ in addition to $\Omega_{\rm GW}(f)$, and (iii) combining all three—$\Omega_{\rm GW}(f)$, $C_{\ell}^{\rm gGW}(f)$, and $C_{\ell}^{\rm GWGW}(f)$. The posterior distributions and the corresponding correlations between parameters are displayed for each case. The constraints are derived for 2000 isotropically distributed pulsars with noise parameters $A_{\rm rd} = 10^{-15}$, $\gamma = -2.5$, $\sigma_{\rm w} = 10^{-7}$ s. The injected parameter values are $\eta = 8$, $\rho = 1$, $\nu = -0.3$, $\alpha = -2.6$, $\lambda = 0.3$, and $\xi = 0$.}

    \label{fig:Corner1}
\end{figure*}

\section{Fisher Forecast \lowercase{{\large for}} constraining the physical parameters}\label{sec:Fisher}

Unlike individually resolved sources, the SGWB represents a stochastic superposition of numerous unresolved binary signals. The power spectral density of SGWB and its anisotropy encode information about the population of the binaries and its cosmic evolution. Given the computational cost of modeling the SGWB signal, we perform a Fisher analysis to show how well the SGWB and the angular power spectrum can constrain the properties of the SMBHB population and its evolution with redshift. The Fisher matrix formalism is a widely used statistical technique for estimating the precision with which model parameters can be determined from observational data \citep{fisher1935logic,tegmark1997karhunen}. This approach is particularly effective in scenarios where the likelihood function is well-approximated by a multivariate Gaussian. The Fisher information matrix is defined as:

\begin{equation} F_{ij} = \Big<\frac{\partial^{2} \mathcal{L}}{\partial \theta_{i} \partial \theta_{j}}\Big>, \label{Fisher} \end{equation}
where $\mathcal{L} = -\ln L$, and $L$ represents the likelihood function. The parameters $\theta_{i}$ and $\theta_{j}$ are the model parameters being estimated. The Fisher matrix quantifies the amount of information that the data provides about these parameters. The diagonal elements of the inverse of the Fisher matrix ($F^{-1}$) describe the information in the data to individual parameter $\theta_i$, while the off-diagonal elements capture the correlations between different parameters. The Cramér-Rao \citep{rao1945information,cramer1946contribution} bound provides a theoretical lower limit on the uncertainty of parameter estimation, given by 

\begin{equation} \Delta \theta_{i} \geq \sqrt{(F^{-1})_{ii}}. \end{equation}
The Cram\'er--Rao bound estimates the minimum error on the parameters, and does not capture the correlation between the parameters, beyond the Gaussian distribution. However, this demonstrates the scope of this technique in discovering the SMBHBs evolution in the Universe from future observations. Our future work will demonstrate the capability of this technique in a Bayesian framework.

\begin{figure*}
    \centering
    \includegraphics[width=0.9\textwidth, height=0.25\textheight]{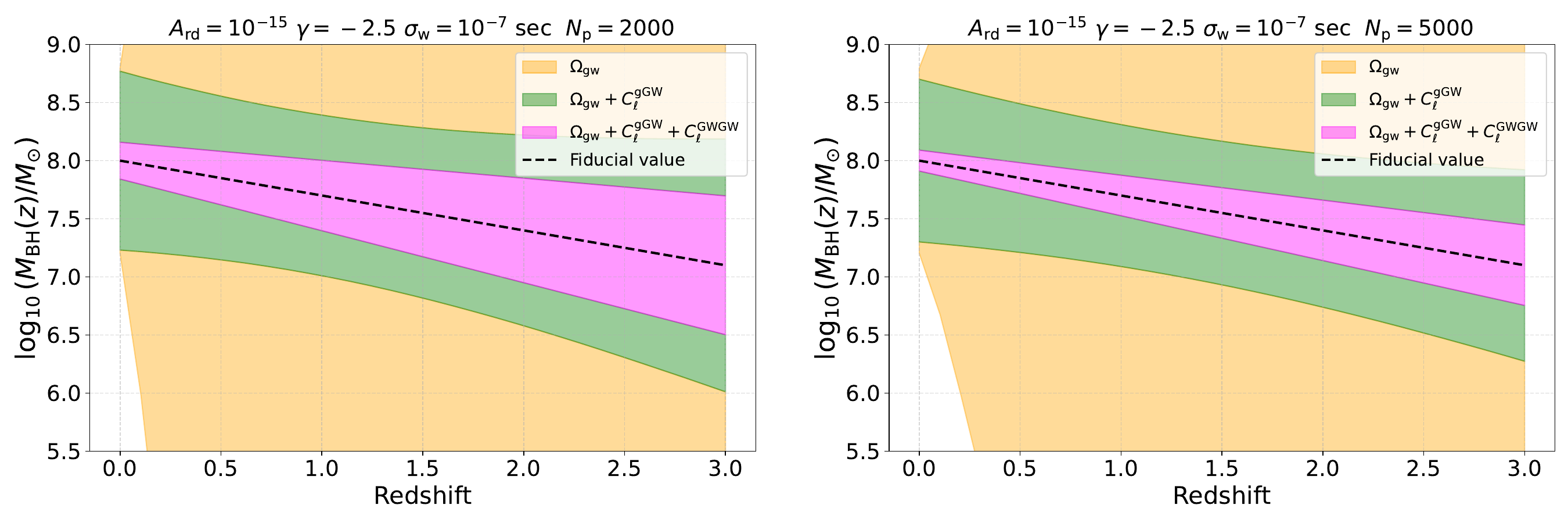}
    \caption{Redshift evolution of the mean mass of the SMBHB, $M_{\rm BH}({\rm z})$, given by Eq.~\eqref{MBH1}, for a stellar mass of $M_{*} = 10^{11} M_{\odot}$. The shaded regions represent the $68\%$ credible interval on $M_{\rm BH}$. The results are shown for two different PTA configurations: (i) $N_{\rm p} = 2000$ (left panel) and (ii) $N_{\rm p} = 5000$ (right panel) pulsars with noise parameters $A_{\rm rd} = 10^{-15}$, $\gamma = -2.5$, and $\sigma_{\rm w} = 10^{-7}$ s. The injected parameter values are $\eta = 8$, $\rho = 1$, $\nu = -0.3$, $\alpha = -2.6$, $\lambda = 0.3$, and $\xi = 0$. A Gaussian prior is applied to $\eta$ and $\rho$, with a standard deviation of $10\%$ of their injected values. The results are presented for three cases: (1) considering only $\Omega_{\rm GW}$, (2) including both $\Omega_{\rm GW}$ and $C_{\ell}^{\rm gGW}$, and (3) incorporating $\Omega_{\rm GW}$, $C_{\ell}^{\rm gGW}$, and $C_{\ell}^{\rm GWGW}$.}
    \label{fig:MBH_10}
\end{figure*}

\begin{figure*}
    \centering
    \includegraphics[width=0.9\textwidth, height=0.25\textheight]{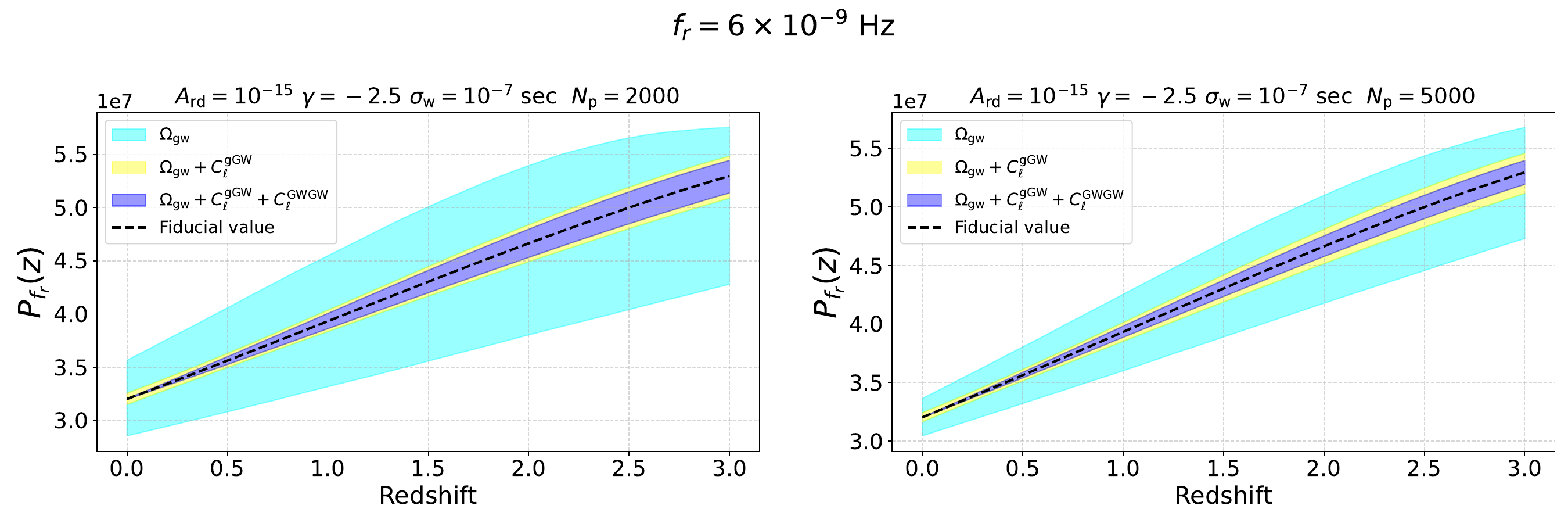}
    
    \caption{Redshift evolution of the normalized frequency distribution, $P_{f_r}({\rm z})$ at $f_r  = 6 \times 10^{-9} ~ \text{Hz}$, of SMBHBs. The shaded regions represent the $68\%$ credible interval.  The results are shown for two different PTA configurations: (i) $N_{\rm p} = 2000$ (left panel) and (ii) $N_{\rm p} = 5000$ (right panel) pulsars with noise parameters $A_{\rm rd} = 10^{-15}$, $\gamma = -2.5$, and  $\sigma_{\rm w} = 10^{-7}$ s. The injected parameter values are $\eta = 8$, $\rho = 1$, $\nu = -0.3$, $\alpha = -2.6$, $\lambda = 0.3$, and $\xi = 0$. A Gaussian prior is applied to $\eta$ and $\rho$, with a standard deviation of $10\%$ of their injected values. The results are presented for three cases: (1) considering only $\Omega_{\rm GW}$, (2) including both $\Omega_{\rm GW}$ and $C_{\ell}^{\rm gGW}$, and (3) incorporating $\Omega_{\rm GW}$, $C_{\ell}^{\rm gGW}$, and $C_{\ell}^{\rm GWGW}$.}   

    \label{fig:dNf1_10}
\end{figure*}

\begin{figure*}
    \centering
    \includegraphics[width=0.9\textwidth, height=0.25\textheight]{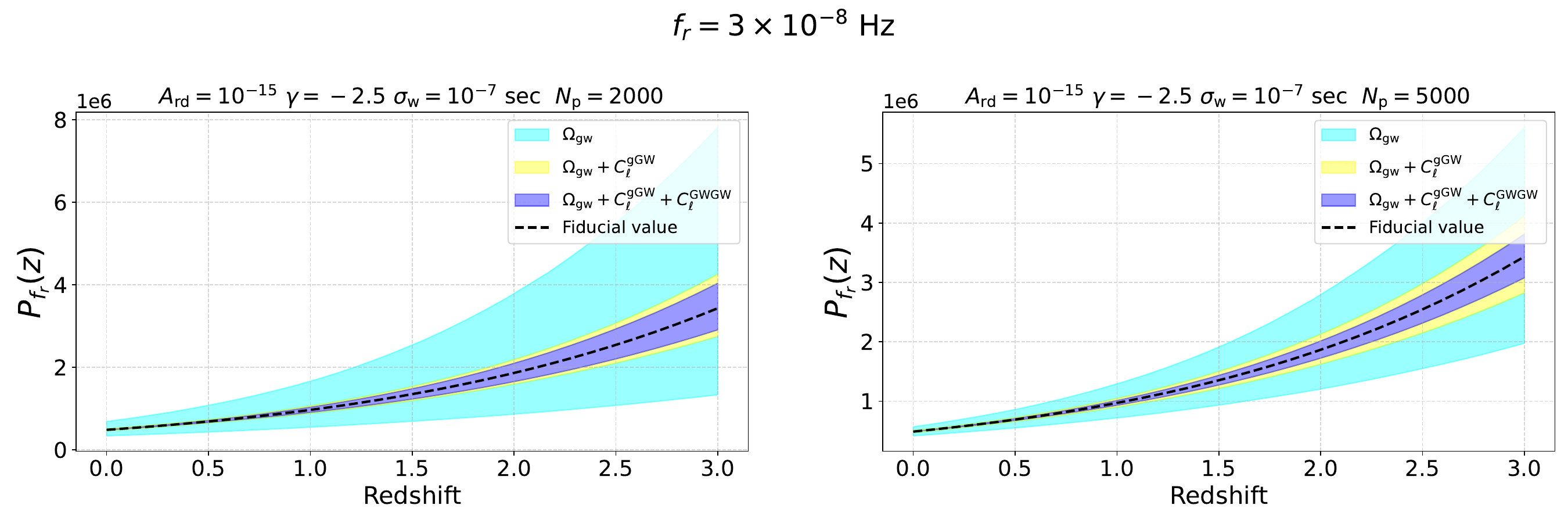}
    
    \caption{Redshift evolution of the normalized frequency distribution, $P_{f_r}({\rm z})$ at $f_r  = 3 \times 10^{-8} ~ \text{Hz}$, of SMBHBs. The shaded regions represent the $68\%$ credible interval.  The results are shown for two different PTA configurations: (i) $N_{\rm p} = 2000$ (left panel) and (ii) $N_{\rm p} = 5000$ (right panel) pulsars with noise parameters $A_{\rm rd} = 10^{-15}$, $\gamma = -2.5$, and  $\sigma_{\rm w} = 10^{-7}$ s. The injected parameter values are $\eta = 8$, $\rho = 1$, $\nu = -0.3$, $\alpha = -2.6$, $\lambda = 0.3$, and $\xi = 0$. A Gaussian prior is applied to $\eta$ and $\rho$, with a standard deviation of $10\%$ of their injected values. The results are presented for three cases: (1) considering only $\Omega_{\rm GW}$, (2) including both $\Omega_{\rm GW}$ and $C_{\ell}^{\rm gGW}$, and (3) incorporating $\Omega_{\rm GW}$, $C_{\ell}^{\rm gGW}$, and $C_{\ell}^{\rm GWGW}$.}     

    \label{fig:dNf2_10}
\end{figure*}

The log-likelihood $\mathcal{L} \equiv -\ln L$ can be written as 
\begin{equation}
    \mathcal{L} = \mathcal{L}_{\rm GW} + \mathcal{L}_{\rm GWGW}+ \mathcal{L}_{\rm gGW}, 
\end{equation}
where
\begin{equation}
    \mathcal{L}_{\rm GW} \propto \sum\limits_{f} \frac{(\hat{\Omega}_{\rm GW}(f) - \Omega_{\rm GW}(f,\Theta))^2 }{2 ~ \Sigma^{2}_{\rm GW}(f,\Theta)},
\end{equation}
where $\mathcal{L}_{\rm GW}$ represents the likelihood of the $\hat{\Omega}_{\rm GW}(f)$ given a set of model parameter $\Theta$, and $\Sigma_{\rm GW}(f,\Theta)$ is the uncertainty in the measurement of the $\Omega_{\rm GW}(f,\Theta)$.
\begin{equation}
    \mathcal{L}_{\rm GWGW} \propto \sum\limits_{f,\ell} \frac{(\hat{C}_{\ell}^{\rm GWGW}(f) - C_{\ell}^{\rm GWGW}(f,\Theta))^2 }{2  ~\Sigma^{2}_{\rm GWGW}(f,\ell,\Theta)},
\end{equation}
where $\mathcal{L}_{\rm GWGW}$ represents the likelihood of the $\hat{C}_{\ell}^{\rm GWGW}(f)$ given a set of model parameter $\Theta$, and $\Sigma_{\rm GWGW}(f,\ell,\Theta)$ is the uncertainty in the measurement of the $C_{\ell}^{\rm GWGW}(f,\Theta)$, and
\begin{equation}
    \mathcal{L}_{\rm gGW} \propto \sum\limits_{f,\ell,{\rm z}} \frac{(\hat{C}_{\ell}^{\rm gGW}(f,{\rm z}) - C_{\ell}^{\rm gGW}(f,{\rm z},\Theta))^2 }{2 ~\Sigma^{2}_{\rm gGW}(f,\ell,{\rm z},\Theta)},
\end{equation}
where $\mathcal{L}_{\rm gGW}$ represents the likelihood of the $\hat{C}_{\ell}^{\rm gGW}(f,{\rm z})$ given a set of model parameter $\Theta$, and $\Sigma_{\rm gGW}(f,\ell,{\rm z},\Theta)$ is the uncertainty in the measurement of the $C_{\ell}^{\rm gGW}(f,{\rm z},\Theta)$.

The analysis assumes Gaussian noise for computational simplicity. In reality, the noise may exhibit non-Gaussian features and future studies will explore this aspect.
We calculate the $\Omega_{\rm GW}(f)$ and $C_{\ell}^{\rm gGW}(f)$ signal using the Eq. \eqref{SGWB2} and Eq. \eqref{ClgGW_1}, respectively. The estimates from simulations match well with the analytical results. Whereas the $C_{\ell}^{\rm GWGW}(f)$ is obtained by averaging the signal over 1000 GW source population realizations using Eq. \eqref{CGWGW}. The $C_{\ell}^{\rm GWGW}$ spectrum is dominated by shot noise arising from the stochastic distribution of a finite number of SMBHB sources, requiring statistical averaging over multiple realizations. Simulations are essential in this context, as they capture the variance due to a finite number of sources that are not adequately described by purely analytical approaches.

The uncertainty in the measurement of these quantities is obtained by using the white and red noise expressions as described in Sec. \ref{sec:Cross}. The uncertainty in the $C_{\ell}^{\rm GWGW}(f)$ and $C_{\ell}^{\rm gGW}(f,\rm{z})$ is given by Eq. \eqref{Var1} and Eq. \eqref{Var2} respectively. The noise angular power spectrum, $N_{\ell}$, is derived in Appendix \ref{sec:appendixA}. In this analysis, we assume an isotropic distribution of pulsars with same noise characteristics for all pulsars. However, the formalism can be easily extended for any pulsar distribution. For this work, we set the noise parameters as $A_{\rm rd} = 10^{-15}$, $\gamma = -2.5$, $\sigma_{\rm w} = 1 \times 10^{-7}~\mathrm{sec}$, a pulse timing cadence of $\Delta t = 30$ days, and a total observation duration of 20 years. These values are motivated by the noise characteristics observed in current PTA experiments as well as expected from SKA-PTA \citep{agazie2023nanograv,wang2017pulsar}. The results with pulsars with higher red noise are shown in the Appendix \ref{sec:appendixB}. A realistic anisotropic distribution of pulsars will introduce additional complexities and decrease the sensitivity to the GW signal. In particular, anisotropy in the pulsar distribution affects the response function of the PTA, leading to a directional dependence in the signal-to-noise ratio across the sky. This can impact the detectability of anisotropic features in the SGWB.

We perform a Fisher analysis for three distinct cases: (1) using $ \Omega_{\rm GW}(f) $ alone, (2) combining $ \Omega_{\rm GW}(f) $ and $ C_{\ell}^{\rm gGW}(f) $, and (3) incorporating $\Omega_{\rm GW}(f)$, $C_{\ell}^{\rm gGW}(f)$, and $C_{\ell}^{\rm GWGW}(f)$. The analysis focuses on constraining the parameters,  $\eta $, $\rho$, $ \nu $, $ \alpha $, $ \lambda $, and $\xi$. The parameters $\mathcal{N}$ quantify the occupation fraction of the SMBHBs in galaxies. 
This parameter is completely degenerate with $\eta$, $\rho$, and $\nu$ as all of these control the overall magnitude of the  $\Omega_{\rm GW}$. We can fix the value to the $\Omega_{\rm GW}(f_{\rm ref})$, where $f_{\rm ref}$ can be the most sensitive frequency.

In Fig.~\ref{fig:Corner1}, we present the results of the Fisher analysis for the three cases under consideration. We assume an array of 2000 isotropically distributed pulsars with identical noise properties. For comparison, in the Square Kilometre Array (SKA) era, PTAs are expected to time nearly 6000 millisecond pulsars \citep{smits2009pulsar}. The injected parameter values are $\eta = 8$, $\rho = 1$, $\nu = -0.3$, $\alpha = -2.6$, $\lambda = 0.3$, and $\xi = 0$ . The contour plots indicate that when only $\Omega_{\rm GW}(f)$ is used, no meaningful constraints can be placed on any parameters due to significant degeneracies, particularly between $\eta$ and $\rho$, $\nu$ and $\xi$. The $ M_{*} - M_{\rm BH} $ relation cannot be directly measured with the SGWB, as it primarily probes the population properties of SMBH binaries rather than individual host galaxy correlations. The SGWB spectrum reflects the binary distribution over cosmic time, and additional constraints, such as the detection of individual GW sources and its host galaxy can help in breaking the degeneracies. The inclusion of the cross-correlation spectrum $C_{\ell}^{\rm gGW}(f)$ improves the constraints on the spectral parameters $\alpha$, $\lambda$, and particularly on $\nu$. When the auto-correlation $C_{\ell}^{\rm GWGW}(f)$ is also incorporated into the analysis, we observe significantly tighter constraints on both $\alpha$ and $\lambda$. The constraint on the redshift evolution parameter for binary formation, $\xi$, also improves with the inclusion of anisotropic signals. Since the parameter governs the redshift contribution of the GW energy to the SGWB, the cross-correlation spectrum is sensitive to this parameter. However, the constraint on $\xi$ remains weak because of its degeneracy with other parameters.

Despite the high signal-to-noise ratio (SNR) with which $\Omega_{\rm GW}(f)$ can be measured, strong parameter degeneracies significantly limit its constraining power. In contrast, 
$C_{\ell}^{\rm GWGW}$ and $C_{\ell}^{\rm gGW}$ individually do not provide any constraints owing to their low SNR, but their combination with $\Omega_{\rm GW}(f)$ helps to break these degeneracies and yields meaningful constraints. This improvement arises from the distinct parameter correlations inherent in $\Omega_{\rm GW}(f)$, $C_{\ell}^{\rm gGW}$, and $C_{\ell}^{\rm GWGW}$. This highlights the crucial role of anisotropy signals in breaking degeneracies related to the redshift evolution of the SMBHB population.

It is worth noting that $C_{\ell}^{\rm gGW}$ is unaffected by changes in the overall magnitude of the signal. However, it is sensitive to the redshift evolution of the GW source, as it depends on the relative contributions of GWs from different redshifts. As we discussed in Sec. \ref{sec:Map} with Fig. \ref{Par_Phy1}, the observables are differently affected by changes in the $\eta$ and $\nu$. While $\eta$ does not affect the features of the $C_{\ell}^{\rm gGW}$, the observables are significantly affected by the change in $\nu$. This sensitivity helps to break the degeneracies that exist between the parameters. The $\alpha$ governs the distribution of SMBHBs emitting at different frequencies, while $\lambda$ controls its redshift evolution. Tight constraints on these parameters imply that PTAs can effectively probe the frequency distribution and its evolution with cosmic time.

In Fig. \ref{fig:MBH_10}, we illustrate the evolution of the SMBHB mass as a function of redshift, along with the corresponding $68 \%$ credible interval on the measurement of the $\rm Log_{10}(M_{\rm BH}({\rm z})/M_{\odot})$, for a PTA with noise parameters $A_{\rm rd}=10^{-15}$, $\gamma=-2.5$, and $\sigma_{\rm w}=10^{-7}\,\mathrm{s}$. Results are presented for arrays of 2000 and 5000 pulsars. The black dashed line indicates the injected values of $\log_{10}(M_{\rm BH}({\rm z})/M_{\odot})$. The injected parameter values are $\eta = 8$, $\rho = 1$, $\nu = -0.3$, $\alpha = -2.6$, $\lambda = 0.3$, and $\xi = 0$.  The quantity $M_{\rm BH}$ represents the mean primary mass of the SMBHB, given by Eq. \eqref{MBH1}, residing in a galaxy with a stellar mass of $M_{*} = 10^{11} M_{\odot}$. The $M_{*}$–$M_{\rm BH}$ relation is modeled based on insights from simulations and observational studies \citep{reines2015relations,habouzit2021supermassive,kozhikkal2024mass}, which examine this correlation for both isolated and binary BHs. The typical values of $\eta$ and $\rho$ are constrained by these studies, with variations estimated to be within $\sim 10\%$. We adopt this as a prior to quantify the uncertainty in the redshift evolution of $M_{\rm BH}$, applying Gaussian priors on both $\eta$ and $\rho$ with standard deviations equal to $10\%$ of their fiducial values. The shaded region in the figure represents the $68\%$ confidence interval.

The results are presented for three cases: (1) considering only $\Omega_{\rm GW}$, (2) including both $\Omega_{\rm GW}$ and $C_{\ell}^{\rm gGW}$, and (3) incorporating $\Omega_{\rm GW}$, $C_{\ell}^{\rm gGW}$, and $C_{\ell}^{\rm GWGW}$. Consistent with the parameter constraints discussed earlier, we find that adding $C_{\ell}^{\rm gGW}$ significantly enhances the measurement of the redshift evolution of $M_{\rm BH}$, while including $C_{\ell}^{\rm GWGW}(f)$ further refines these constraints. Our results indicate that, for the noise configuration considered here, the mean SMBH 
mass, $M_{\rm BH}$, associated with a host galaxy of fixed stellar mass can be measured 
with a signal-to-noise ratio 
($\mathrm{SNR}= M_{\rm BH}({\rm z})/\sigma_{M_{\rm BH}}({\rm z})$) 
of $\sim 1.8$ at ${\rm z}=0.5$ and $\sim 0.8$ at ${\rm z}=2$ for an array of 2000 pulsars, 
and $\sim 3$ at ${\rm z}=0.5$ and $\sim 1.5$ at ${\rm z}=2$ for 5000 pulsars.

\begin{table*}
    \centering
    \renewcommand{\arraystretch}{1.5}  
    \setlength{\tabcolsep}{8pt}       
    \begin{tabular}{|c|c|c|}
        \hline
        & \multicolumn{2}{c|}{Ratio of Figure of Merit (FoM) relative to $\Omega_{\rm GW}$} \\
        \hline
        SGWB signals used for the analysis 
        & $\qquad \quad  N_{p}=2000 \qquad \quad $ 
        & $N_{p}=5000$ \\  
        \hline
        $\Omega_{\rm GW} + C_{\ell}^{\rm gGW}$ 
        & 21.6 & 17.4 \\  
        \hline
        $\Omega_{\rm GW} + C_{\ell}^{\rm gGW} + C_{\ell}^{\rm GWGW}$ 
        & 221 & 254 \\  
        \hline
    \end{tabular}
    \caption{Ratios of the Figure of Merit (FoM) for different combinations of SGWB signals relative to the case with only $\Omega_{\rm GW}$, shown for different numbers of isotropically distributed pulsars with noise parameters $A_{\rm rd} = 10^{-15}$, $\gamma = -2.5$, and $\sigma_{\rm w} = 10^{-7}$ s.}
    \label{table}
\end{table*}

Similarly, Fig.~\ref{fig:dNf1_10} and Fig.~\ref{fig:dNf2_10}, we present the redshift evolution of the normalized frequency distribution, represented as $P_{f_r}$, under similar conditions as Fig. \ref{fig:MBH_10}, at $f_r = 6 \times 10^{-9}~\text{Hz}$ and $f_r = 3 \times 10^{-8}~\text{Hz}$, respectively. As expected, the constraints on $\alpha$ and $\lambda$ lead to an increasingly precise determination of the redshift dependence of the GW source properties when anisotropy observables are included. In both cases, the improvement is more pronounced when increasing the number of pulsars from 2000 to 5000, further demonstrating the potential of PTAs to probe the SMBHB population with higher precision. The results show that, for the noise configuration considered here, the frequency distribution can be measured with a signal-to-noise ratio ($\mathrm{SNR}=P_{f_r}({\rm z})\big/\sigma_{P_{f_r}}$)
ranging from $\sim 10$ to $\sim 100$ between ${\rm z}=0.5$ and ${\rm z}=2$ for an array of 2000 pulsars, and from $\sim 20$ to $\sim 150$ over the same redshift range for 5000 pulsars.

In Table \ref{table}, we show the ratio of the Figure of Merit (FoM) for two cases (i) $\Omega_{\rm GW}$ combined with $C_{\ell}^{\rm gGW}$ and (ii) the combination of $\Omega_{\rm GW}$, $C_{\ell}^{\rm gGW}$, and $C_{\ell}^{\rm GWGW}$, relative to the case with only $\Omega_{\rm GW}$. The FoM, defined as the square root of the determinant of the Fisher matrix, quantifies the information content and effectiveness of parameter estimation. The table compares the FoM ratio for different noise configurations and different numbers of isotropically distributed pulsars. 
From the table, we observe that adding the  $C_{\ell}^{\rm gGW}$ to $\Omega_{\rm GW}$ enhances the FoM by approximately a factor of 20, depending on the number of pulsars included in the array. Furthermore, incorporating the SGWB auto-correlation term ($C_{\ell}^{\rm GWGW}$) leads to a much more significant enhancement in FoM, with values increasing up to a factor of 250. This demonstrates the substantial gain in information when combining complementary information from $\Omega_{\rm GW}$, $C_{\ell}^{\rm gGW}$, and $C_{\ell}^{\rm GWGW}$.


These results demonstrate that next-generation PTAs, such as SKA with its planned large number of pulsars, will be capable of providing meaningful constraints on the SMBHB population and its redshift evolution. Specifically, our Fisher analysis reveals that while the spectral energy density, $\Omega_{\rm GW}(f)$, alone suffers from strong parameter degeneracies, the inclusion of anisotropic observables such as the angular power spectrum, $C_{\ell}^{\rm GWGW}(f)$, and the cross-angular power spectrum with galaxy surveys, $C_{\ell}^{\rm gGW}(f)$, significantly improves parameter estimation. 

These observables allow us to improve the measurement of key physical parameters, such as the redshift evolution of the SMBHB-Galaxy relation and the frequency distribution of the binaries, thereby refining our understanding of their cosmic history. Notably, our analysis indicates that constraints on the redshift evolution of SMBHBs are particularly sensitive to cross-correlation measurements, highlighting the crucial role of galaxy surveys in breaking degeneracies. By integrating future PTA datasets, such as those from SKA, with large-scale structure surveys, these constraints can be further refined, providing deeper insights into the formation and growth of SMBHs across cosmic time.

While our analysis demonstrates the potential of PTAs in constraining the SMBHB population, several caveats must be considered. First, the sensitivity of PTAs is inherently limited by the number of pulsars, their spatial distribution, and the precision of timing residual measurements. Any unmodeled noise sources, such as intrinsic pulsar spin irregularities, could introduce biases in parameter estimation. Our analysis further assumes an idealized PTA with isotropically distributed pulsars having uniform noise properties, whereas real PTAs exhibit anisotropic sky coverage and varying timing precision. However, this anisotropy in pulsar distribution does not present a fundamental limitation; it primarily introduces direction-dependent noise, which affects the angular resolution rather than the detectability of the signal. 

We also neglect the contribution from cosmological SGWB sources such as cosmic inflation, cosmic strings, and phase transitions \citep{ellis2024source,madge2023primordial}. The inclusion of such cosmological components would only modify the overall $\Omega_{\rm GW}(f)$ spectrum, since the primordial SGWB is expected to be significantly less anisotropic compared to the background from a population of SMBHB \citep{gardiner2024beyond,sato2024exploring, sah2024imprints,valbusa2021imprint,contaldi2017anisotropies}. The presence of a cosmological component can be straightforwardly accommodated by introducing additional parameters describing its amplitude and spectral shape. 
Furthermore, including non-zero orbital eccentricity introduces spectral correlations between frequencies in the SGWB, as power is emitted over multiple harmonics of the orbital frequency \citep{sah2025accurate}. These correlations encode information about the eccentricity distribution of the SMBHB population, as demonstrated in  \cite{sah2025accurate}. The eccentricity can also modify the overall SGWB spectrum and can introduce degeneracies with environmental effects that influence binary evolution. However, the induced spectral correlations can help break this degeneracy by providing an additional observable signature. The anisotropy, on the other hand, remains largely unaffected by eccentricity, since it is determined mainly by the spatial distribution of the sources. We plan to explore these effects in future work.

Future improvements will focus on incorporating realistic pulsar distributions and spatially varying noise models to refine our predictions. Additionally, deeper galaxy surveys with spectroscopic redshifts, such as Rubin LSST \citep{ivezic2019lsst}, DESI \citep{flaugher2014dark,levi2013desi} and Euclid \citep{laureijs2010euclid,racca2016euclid}, will significantly enhance the cross-correlation analysis by providing more precise three-dimensional maps of the galaxy distribution. This will allow for a more accurate determination of the redshift evolution of the SMBHB population. The combination of improved PTAs and more detailed galaxy surveys will ultimately enable a more robust understanding of the connection between SMBHBs and large-scale structures.

\section{Conclusion}  \label{sec:Conc}

The nHz GW provides a unique probe for understanding the formation and evolution of SMBHBs. The detection of the SGWB by PTAs offers an opportunity to study the unresolved SMBHB population and its connection to galaxy evolution. In this work, we have demonstrated the feasibility of using the SGWB and its anisotropy to constrain the population and evolution of SMBHBs. By analyzing the spectral density of the SGWB, $\Omega_{\rm GW}(f)$, its angular power spectrum, $C_{\ell}^{\rm GWGW}(f)$, and its cross-angular power spectrum with galaxy density fluctuations, $C_{\ell}^{\rm gGW}(f)$, we have explored the ability of PTAs to probe the SMBHB population properties and their redshift evolution. We show that anisotropic SGWB measurements significantly improve parameter estimation and break degeneracies in SMBHB population modeling. 

Our Fisher analysis shows that $\Omega_{\rm GW}(f)$ alone offers limited constraining power due to parameter degeneracies, particularly between the SMBHB mass distribution parameters ($\eta$ and $\rho$) and its redshift evolution ($\nu$). However, incorporating $C_{\ell}^{\rm GWGW}(f)$ and $C_{\ell}^{\rm gGW}(f)$ substantially improves these constraints, with $C_{\ell}^{\rm gGW}(f)$ being especially effective in breaking degeneracies linked to the redshift evolution of the SMBHB mass function. To further address the degeneracy between $\eta$ and $\rho$, we impose $10\%$ Gaussian priors on both, motivated by local observations of the $M_{*}-M_{\rm BH}$ relation. Assuming PTA configurations with 2000 and 5000 pulsars, feasible in the SKA era, we find that the combined use of isotropic and anisotropic SGWB signals facilitates meaningful measurements of the SMBHB mass evolution and its frequency distribution across cosmic time. These results highlight the potential of future PTAs to probe the growth and evolution of the SMBHB population with significant detail.

Our results establish a benchmark for future studies of SMBHB populations using PTAs and galaxy surveys. Next-generation PTAs, such as SKA, with their increased number of pulsars and improved noise sensitivity, will provide better constraints on the SMBHB population. Additionally, future galaxy surveys like Rubin LSST, with improved redshift accuracy, will refine the cross-correlation analysis. Incorporating multi-frequency GW observations from space-based detectors like LISA will further complement PTA measurements by probing SMBHBs across a broader mass and redshift range. These advancements will provide a more comprehensive picture of the formation and evolution of SMBHBs.

\section*{Acknowledgments}
This work is part of the $\langle \texttt{data|theory}\rangle$ \texttt{Universe-Lab} which is supported by the TIFR and the Department of Atomic Energy, Government of India. The authors would like to thank the $\langle \texttt{data|theory}\rangle$ \texttt{Universe-Lab} for providing computing resources. The author thanks Chiara Mingarelli and Federico Semenzato for their valuable feedback during NANOGrav paper circulation. The authors would also like to acknowledge the use of the following Python packages in this work: Numpy \citep{van2011numpy,2020NumPy-Array}, Scipy \citep{jones2001scipy,virtanen2020scipy}, Matplotlib \citep{hunter2007matplotlib}, Astropy \citep{robitaille2013astropy,price2018astropy,2022ApJ...935..167A}, Hasasia \citep{Hazboun2019Hasasia}, Ray \citep{moritz2018ray}, Pygtc \citep{bocquet2019pygtc}, and PTArcade \citep{mitridate2023ptarcade,lamb2023need}.

\bibliographystyle{aasjournal}
\bibliography{main}

\label{lastpage}

\appendix
\section{Derivation \lowercase{{\large of}} the noise angular power spectrum}
\label{sec:appendixA}

The cross-correlation of Fourier transform of the timing residuals from two pulsars I and J is given by:

\begin{align}
    \langle \tilde{r}_I(f) \tilde{r}_J^*(f) \rangle = B(f) \times & \int d\hat{n} ~ \Omega_{\rm GW}(f,\hat{n}) \\ & \nonumber \Big(\mathcal{F}^{+}_{I}(\hat{n}) \mathcal{F}^{+}_{J}(\hat{n}) + \mathcal{F}^{\times}_{I}(\hat{n}) \mathcal{F}^{\times}_{J}(\hat{n})\Big),
\end{align}
where $B(f)= 4\pi \frac{3 H_0^2}{8 \pi^4} \frac{f^{-5}}{\Delta f}$ and 
\begin{equation}
     \mathcal{F}^{A}_{I}(\hat{n}) = \frac{1}{2} \frac{\hat{p_I}^{a}\hat{p_I}^{b}}{1+\hat{n}  \cdot \hat{p_I}} e^{A}_{ab}.
\end{equation}

The overlap reduction function between pulsars I and J, $\Gamma_{IJ}$, in an equal-pixel basis, called radiometer basis \citep{mitra2008gravitational,thrane2009probing} is given by

\begin{equation}
    \Gamma_{IJ}(f) = \sum\limits_{k} ~\Omega_{\rm GW}(f,\hat{n}_k) \Big(\mathcal{F}^{+}_{I}(\hat{n}_k) \mathcal{F}^{+}_{J}(\hat{n}_k) + \mathcal{F}^{\times}_{I}(\hat{n}_k) \mathcal{F}^{\times}_{J}(\hat{n}_k)\Big).
\end{equation}
We can represent this in a matrix form as 
\begin{equation}
    \mathbf{\Gamma} = \textbf{R} ~ \mathbf{\Omega},
\end{equation}
where $\textbf{R}^{IJ}_{k}  \equiv  \sum\limits_A \mathcal{F}^{A}_{I}(\hat{n}_k)  \mathcal{F}^{A}_{J}(\hat{n}_k)$ and $\mathbf{\Omega_{k}} \equiv \Omega_{\rm GW}(f,\hat{n}_k)$

\begin{equation}
    \mathcal{P}(\hat{\textbf{X}}|\mathbf{\Omega}) \propto \textbf{exp}[\frac{-1}{2} (\hat{\textbf{X}}(f) - \textbf{R} \mathbf{\Omega}(f))^{T} \Sigma^{-1}(f) (\hat{\textbf{X}}(f) - \textbf{R} \mathbf{\Omega}(f))],
\end{equation}
where $\hat{\textbf{X}}_{IJ}$(f) = $\tilde{r}_I(f) \tilde{r}_J^*(f)/B(f)$, $\Sigma(f)$ is the cross-correlation covariance matrix.

The uncertainty in the measurement of $\mathbf{\Omega_{k}}$ denoted by $\mathbf{\Sigma}_{\rm pix}^2$ is given by the inverse of the Fisher matrix $\textbf{R}^{T} \bf{\Sigma}^{-1} \textbf{R}$.

For an isotropic distribution of pulsars with the same pulsar timing residual noise for every pulsar, the noise angular power spectrum case can be written as 

\begin{equation}
    \begin{aligned}
       N_{\ell}  = & \frac{4\pi}{N_{\rm pix}} \sigma^2_{\rm pix},
    \end{aligned}
\end{equation}
where, $\sigma_{\rm pix}$ is standard deviation of pixel noise.

\section{Constraint \lowercase{{\large on}} the evolution \lowercase{{\large of}} the SMBHB population}
\label{sec:appendixB}

\begin{figure*}
    \centering
    \includegraphics[width=0.9\textwidth, height=0.25\textheight]{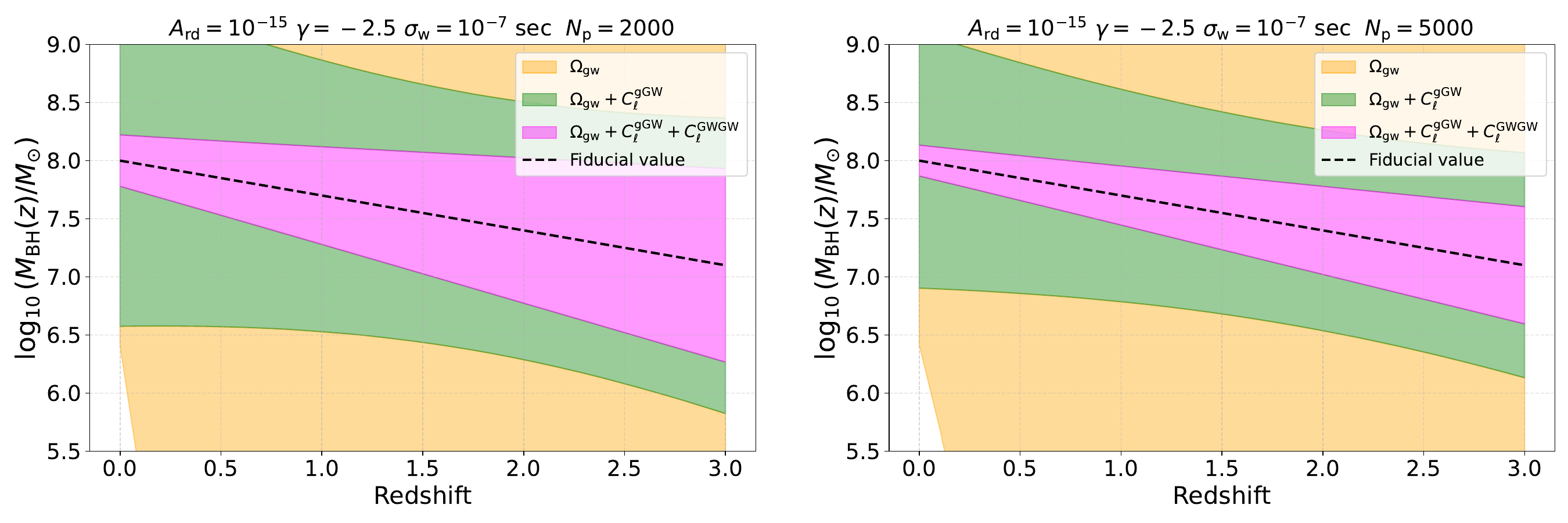}
  
    \caption{Redshift evolution of the mean mass of the SMBHB, $M_{\rm BH}({\rm z})$, given by Eq.~\eqref{MBH1}, for a stellar mass of $M_{*} = 10^{11} M_{\odot}$. The shaded regions represent the $68\%$ credible interval on $M_{\rm BH}$. The results are shown for two different PTA configurations: (i) $N_{\rm p} = 2000$ (left panel) and (ii) $N_{\rm p} = 5000$ (right panel) pulsars with noise parameters $A_{\rm rd} = 10^{-15}$, $\gamma = -2.5$, and $\sigma_{\rm w} = 10^{-7}$ s. The injected parameter values are $\eta = 8$, $\rho = 1$, $\nu = -0.3$, $\alpha = -2.6$, $\lambda = 0.3$, and $\xi = 0$. A Gaussian prior is applied to $\eta$ and $\rho$, with a standard deviation of $20\%$ of their injected values. The results are presented for three cases: (1) considering only $\Omega_{\rm GW}$, (2) including both $\Omega_{\rm GW}$ and $C_{\ell}^{\rm gGW}$, and (3) incorporating $\Omega_{\rm GW}$, $C_{\ell}^{\rm gGW}$, and $C_{\ell}^{\rm GWGW}$.}  

    \label{fig:A_MBH_20}
\end{figure*}

We present the results obtained in Sec.~\ref{sec:Fisher} for the redshift evolution of the SMBHB population when adopting broader priors on $\eta$ and $\rho$. Specifically, we assume Gaussian priors with standard deviations equal to $20\%$ of their fiducial values. Figure~\ref{fig:A_MBH_20} shows the evolution of the SMBHB mass as a function of redshift, together with the associated uncertainties, for a PTA with noise parameters $A_{\rm rd}=10^{-15}$, $\gamma=-2.5$, and $\sigma_{\rm w}=10^{-7}\,\mathrm{s}$. Results are shown for arrays of 2000 and 5000 pulsars. Our analysis indicates that, under these broader priors, the mean SMBH mass ($M_{\rm BH}({\rm z})$) in galaxies of fixed stellar mass can still be constrained, though with weaker precision compared to the case with tighter priors. The mean SMBH mass ($M_{\rm BH}$) residing in the galaxy with a given stellar mass can be detected with an SNR  ($M_{\rm BH}({\rm z})/\sigma_{M_{\rm BH}}({\rm z})$) of $\sim 1.2$ at ${\rm z}=0.5$ and $\sim 0.5$ at ${\rm z}=2$ for an array of 2000 pulsars, 
and $\sim 2$ at ${\rm z}=0.5$ and $\sim 1$ at ${\rm z}=2$ for 5000 pulsars.

\begin{figure*}
    \centering
    \includegraphics[width=0.9\textwidth, height=0.25\textheight]{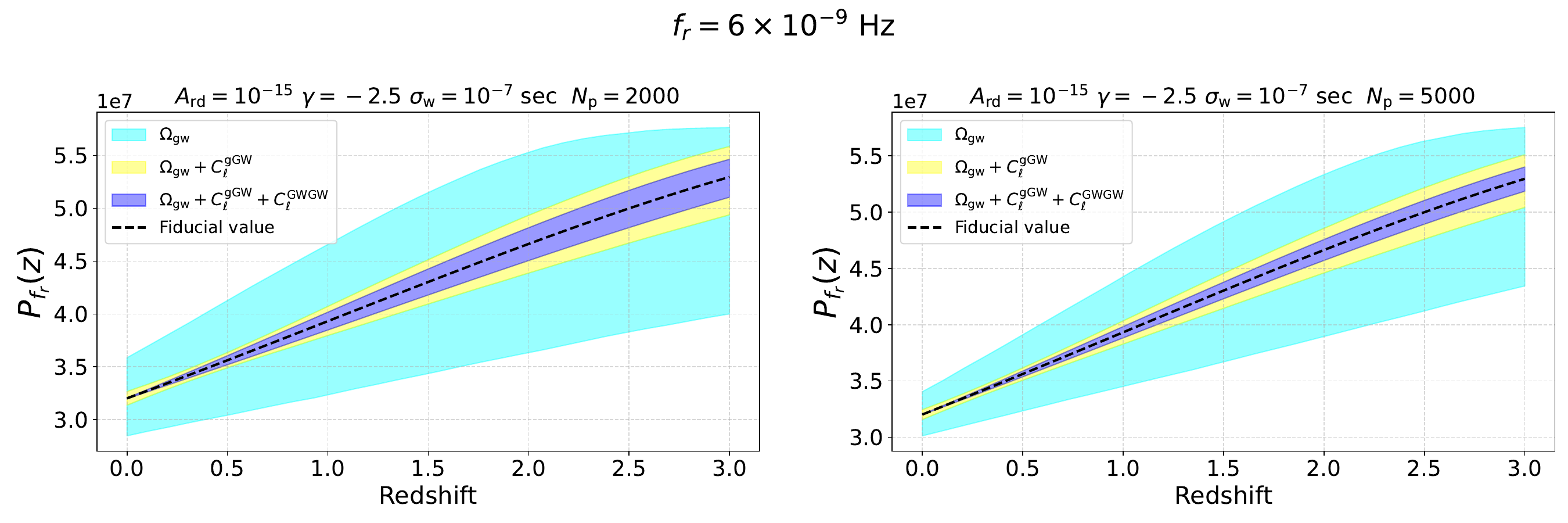}
    
    \caption{Redshift evolution of the normalized frequency distribution, $P_{f_r}({\rm z})$ at $f_r  = 6 \times 10^{-9} ~ \text{Hz}$, of SMBHBs. The shaded regions represent the $68\%$ credible interval.  The results are shown for two different PTA configurations: (i) $N_{\rm p} = 2000$ (left panel) and (ii) $N_{\rm p} = 5000$ (right panel) pulsars with noise parameters $A_{\rm rd} = 10^{-15}$, $\gamma = -2.5$, and $\sigma_{\rm w} = 10^{-7}$ s. The injected parameter values are $\eta = 8$, $\rho = 1$, $\nu = -0.3$, $\alpha = -2.6$, $\lambda = 0.3$, and $\xi = 0$. A Gaussian prior is applied to $\eta$ and $\rho$, with a standard deviation of $20\%$ of their injected values. The results are presented for three cases: (1) considering only $\Omega_{\rm GW}$, (2) including both $\Omega_{\rm GW}$ and $C_{\ell}^{\rm gGW}$, and (3) incorporating $\Omega_{\rm GW}$, $C_{\ell}^{\rm gGW}$, and $C_{\ell}^{\rm GWGW}$.}      

    \label{fig:A_dNf1_20}
\end{figure*}

\begin{figure*}
    \centering
    \includegraphics[width=0.9\textwidth, height=0.25\textheight]{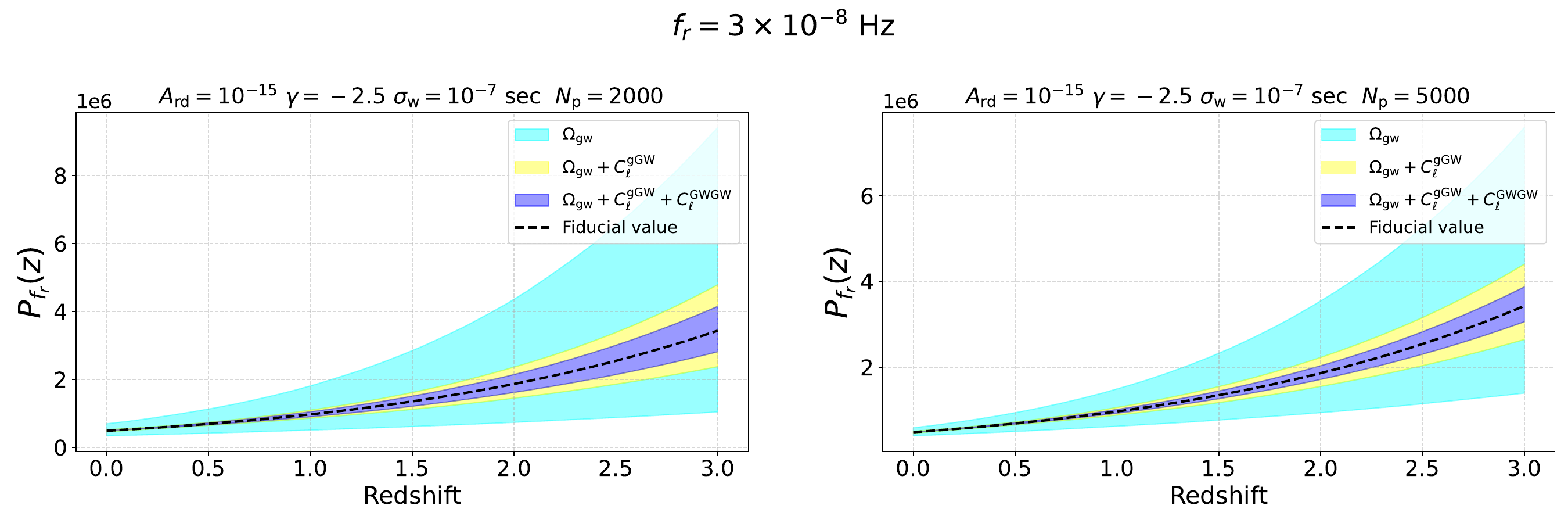}
    
    \caption{Redshift evolution of the normalized frequency distribution, $P_{f_r}({\rm z})$  at $f_r  = 3 \times 10^{-8} ~ \text{Hz}$, of SMBHBs. The shaded regions represent the $68\%$ credible interval. The results are shown for two different PTA configurations: (i) $N_{\rm p} = 2000$ (left panel) and (ii) $N_{\rm p} = 5000$ (right panel) pulsars with noise parameters $A_{\rm rd} = 10^{-15}$, $\gamma = -2.5$, and  $\sigma_{\rm w} = 10^{-7}$ s. The injected parameter values are $\eta = 8$, $\rho = 1$, $\nu = -0.3$, $\alpha = -2.6$, $\lambda = 0.3$, and $\xi = 0$. A Gaussian prior is applied to $\eta$ and $\rho$, with a standard deviation of $20\%$ of their injected values. The results are presented for three cases: (1) considering only $\Omega_{\rm GW}$, (2) including both $\Omega_{\rm GW}$ and $C_{\ell}^{\rm gGW}$, and (3) incorporating $\Omega_{\rm GW}$, $C_{\ell}^{\rm gGW}$, and $C_{\ell}^{\rm GWGW}$.}     

    \label{fig:A_dNf2_20}
\end{figure*}

Similarly, Figs.~\ref{fig:A_dNf1_20} and \ref{fig:A_dNf2_20} present the redshift evolution of the frequency distribution, represented by $P_{f_r}({\rm z})$, under the same conditions. We find that the constraints on the frequency distribution remain nearly unchanged relative to the $10\%$ prior case, since the parameters governing $P_{f_r}({\rm z})$, namely $\alpha$ and $\lambda$, are not strongly degenerate with $\eta$ and $\rho$.

\begin{figure*}
    \centering
    \includegraphics[width=0.9\textwidth, height=0.25\textheight]{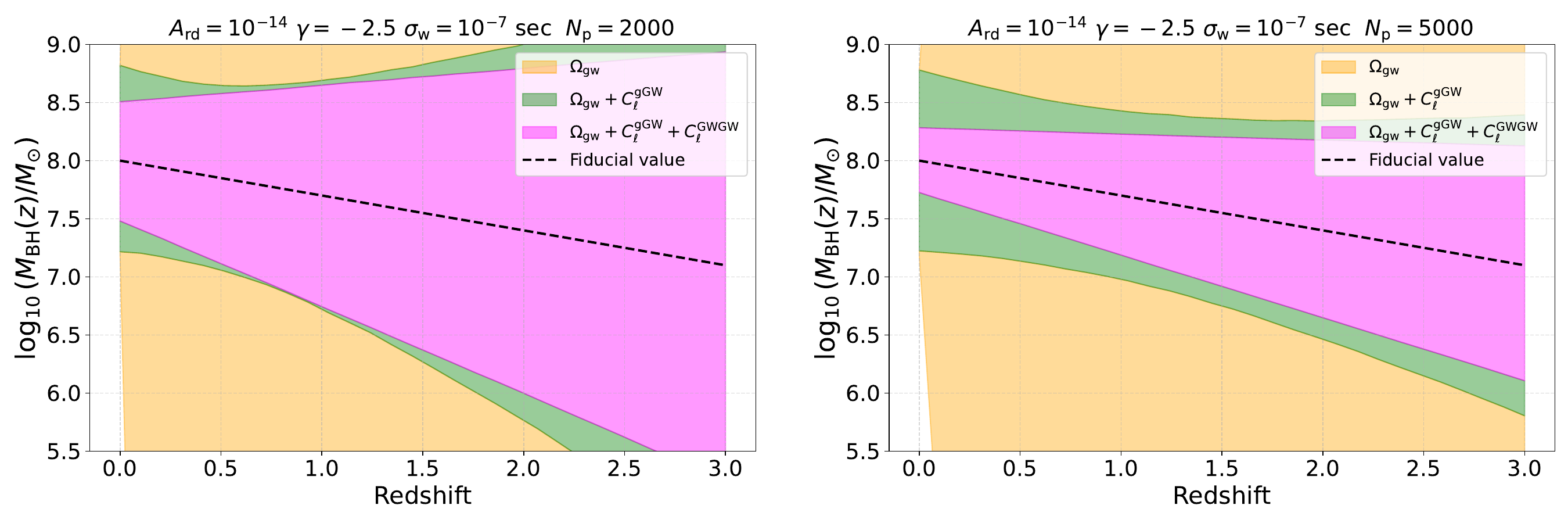}
    \caption{Redshift evolution of the mean mass of the SMBHB, $M_{\rm BH}({\rm z})$, given by Eq.~\eqref{MBH1}, for a stellar mass of $M_{*} = 10^{11} M_{\odot}$ as a function of redshift. The shaded regions represent the $68\%$ credible interval on $M_{\rm BH}$. The results are shown for two different PTA configurations: (i) $N_{\rm p} = 2000$ (left panel) and (ii) $N_{\rm p} = 5000$ (right panel) pulsars with noise parameters $A_{\rm rd} = 10^{-14}$, $\gamma = -2.5$, and $\sigma_{\rm w} = 10^{-7}$ s. The injected parameter values are $\eta = 8$, $\rho = 1$, $\nu = -0.3$, $\alpha = -2.6$, $\lambda = 0.3$, and $\xi = 0$. A Gaussian prior is applied to $\eta$ and $\rho$, with a standard deviation of $10\%$ of their injected values. The results are presented for three cases: (1) considering only $\Omega_{\rm GW}$, (2) including both $\Omega_{\rm GW}$ and $C_{\ell}^{\rm gGW}$, and (3) incorporating $\Omega_{\rm GW}$, $C_{\ell}^{\rm gGW}$, and $C_{\ell}^{\rm GWGW}$.}  

    \label{fig:A_MBH_10-14}
\end{figure*}

\begin{figure*}
    \centering
    \includegraphics[width=0.9\textwidth, height=0.25\textheight]{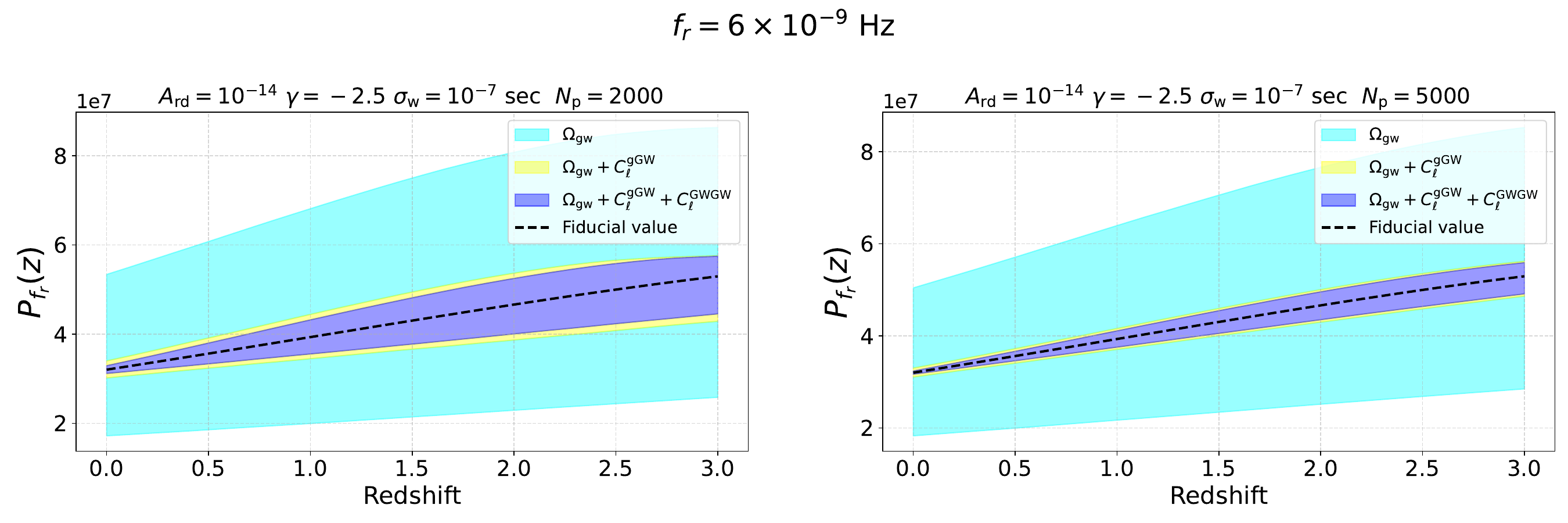}
    
    \caption{Redshift evolution of the normalized frequency distribution, $P_{f_r}({\rm z})$ at $f_r  = 6 \times 10^{-9} ~ \text{Hz}$, of SMBHBs. The shaded regions represent the $68\%$ credible interval.  The results are shown for two different PTA configurations: (i) $N_{\rm p} = 2000$ (left panel) and (ii) $N_{\rm p} = 5000$ (right panel) pulsars with noise parameters $A_{\rm rd} = 10^{-14}$, $\gamma = -2.5$, and $\sigma_{\rm w} = 10^{-7}$ s. The injected parameter values are $\eta = 8$, $\rho = 1$, $\nu = -0.3$, $\alpha = -2.6$, $\lambda = 0.3$, and $\xi = 0$. A Gaussian prior is applied to $\eta$ and $\rho$, with a standard deviation of $10\%$ of their injected values. The results are presented for three cases: (1) considering only $\Omega_{\rm GW}$, (2) including both $\Omega_{\rm GW}$ and $C_{\ell}^{\rm gGW}$, and (3) incorporating $\Omega_{\rm GW}$, $C_{\ell}^{\rm gGW}$, and $C_{\ell}^{\rm GWGW}$.}      

    \label{fig:A_dNf1_10-14}
\end{figure*}

\begin{figure*}
    \centering
    \includegraphics[width=0.9\textwidth, height=0.25\textheight]{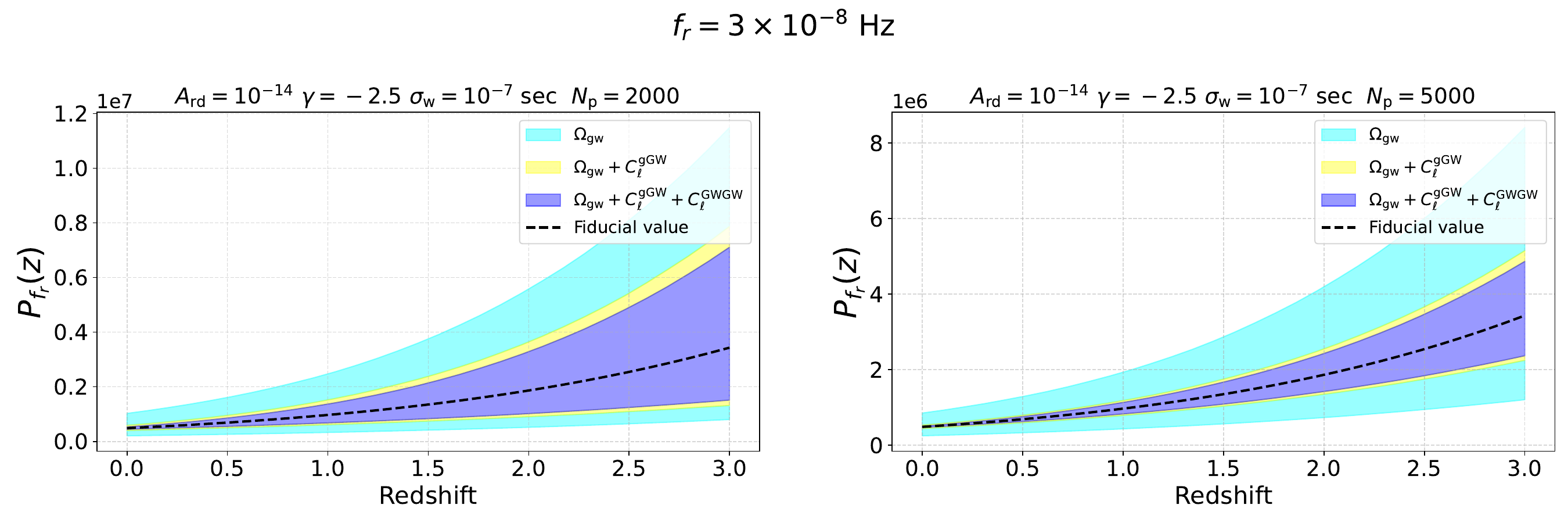}
    
    \caption{Redshift evolution of the normalized frequency distribution, $P_{f_r}({\rm z})$ at $f_r  = 3 \times 10^{-8} ~ \text{Hz}$, of SMBHBs. The shaded regions represent the $68\%$ credible interval. The results are shown for two different PTA configurations: (i) $N_{\rm p} = 2000$ (left panel) and (ii) $N_{\rm p} = 5000$ (right panel) pulsars with noise parameters $A_{\rm rd} = 10^{-14}$, $\gamma = -2.5$, and $\sigma_{\rm w} = 10^{-7}$ s. The injected parameter values are $\eta = 8$, $\rho = 1$, $\nu = -0.3$, $\alpha = -2.6$, $\lambda = 0.3$, and $\xi = 0$. A Gaussian prior is applied to $\eta$ and $\rho$, with a standard deviation of $10\%$ of their injected values. The results are presented for three cases: (1) considering only $\Omega_{\rm GW}$, (2) including both $\Omega_{\rm GW}$ and $C_{\ell}^{\rm gGW}$, and (3) incorporating $\Omega_{\rm GW}$, $C_{\ell}^{\rm gGW}$, and $C_{\ell}^{\rm GWGW}$.}     

    \label{fig:A_dNf2_10-14}
\end{figure*}

We also examine the case with a red noise amplitude an order of magnitude larger ($A_{\rm rd} = 10^{-14}$), adopting Gaussian priors at the $10\%$ level on $\eta$ and $\rho$, as shown in Figs.~\ref{fig:A_MBH_10-14}, \ref{fig:A_dNf1_10-14}, and \ref{fig:A_dNf2_10-14}. In this scenario, the constraints on the redshift evolution of the population become significantly weaker, reflecting the reduced sensitivity of the PTA configuration to the SGWB signal when the red noise level is high.

\end{document}